\documentclass[aps,prb,twocolumn,groupedaddress,showpacs,showkeys]{revtex4-1}
\usepackage{amsmath}
\usepackage{amssymb}
\usepackage{graphicx}
\usepackage{epstopdf}
\usepackage{esint}
\usepackage{tikz}

\begin{document}


\title{Theory  of hyperbolic stratified nanostructures for surface enhanced Raman scattering}

\author{Herman M.K. Wong$^1$, Mohsen Kamandar Dezfouli$^2$, Simon Axelrod$^2$, Stephen Hughes$^2$, and Amr S. Helmy$^1$}
\email{a.helmy@utoronto.ca}

\affiliation{$^1$Department of Electrical and Computer Engineering, University of Toronto, Toronto, ON M5S 3G4, Canada}
\affiliation{$^2$Department of Physics, Engineering Physics and Astronomy, Queen's University, Kingston, ON K7L 3N6, Canada}

\date{\today}

\begin{abstract}
We theoretically investigate the enhancement of surface enhanced Raman spectroscopy (SERS) using hyperbolic stratified nanostructures and compare to metal nanoresonators. The photon Green function of each nanostructure within its environment is first obtained from a semi-analytical modal theory, which is used in a quantum optics formalism of the molecule-nanostructure interaction to model the SERS spectrum. An intuitive methodology is presented for calculating the single molecule enhancement factor (SMEF), which is also able to predict known experimental SERS enhancement factors of an example gold nano-dimer. We elucidate the important figures-of-merit of the enhancement and explore these for different designs. We find that the use of hyperbolic stratified materials can enhance the photonic local density of states (LDOS) by close to 2 times in comparison to pure metal nanostructures, when both designed to work at the same operating wavelengths. However, the increased LDOS is accompanied by higher electric field concentration within the lossy hyperbolic material, which leads to increased quenching that serves to reduce the overall detected SERS enhancement in the far field. For nanoresonators with resonant localized surface plasmon wavelengths in the near-infrared, the SMEF for the hyperbolic stratified nanostructure is approximately an order of magnitude lower than the pure metal counterpart. Conversely, we show that by detecting the Raman signal using a near-field probe, hyperbolic materials can provide an improvement in SERS enhancement compared to using pure metal nanostructures when the probe is sufficiently close ($<$50 nm) to the Raman active molecule at the plasmonic hotspot.

\end{abstract}


\maketitle

\section{Introduction}
\label{Sec1}
Raman spectroscopy has emerged as one of the most commonly utilized techniques for chemical analysis \cite{Baraldi2008,Patel2010,Efremov2008,Downes2010,Tanaka2006,Chalmers2012,Sarmiento2008}. However, spontaneous Raman scattering is exceedingly weak \cite{Kneipp2006}, and a major challenge  is how to separate the Raman scattered photons from the intense Rayleigh scattered pump photons. Indeed, Raman scattering cross-sections are on the order of $10^{-29}$ $\textnormal{cm}^2$, while the cross-sections of typical fluorescence processes are on the order of $10^{-19}$ $\textnormal{cm}^2$\,\,\cite{Kneipp2006}. A formidable technique to boost the signal of Raman scattered light is surface enhanced Raman spectroscopy (SERS), which typically makes use of metals to enhance local electromagnetic (EM) fields. Enhanced EM fields created at a hotspot increases the light-matter interaction of the molecule and the EM field, while simultaneously increasing the Raman scattering rate through the enhancement of the photonic local density of states (LDOS). The combination of field enhancement in both the excitation and the scattering processes helps to enhance the overall collected Raman signal. A commonly accepted approximation entails that the SERS enhancement factor scales with $(|{\bf E}_{\textnormal{loc}}|/|{\bf E}_0|)^4$\,\,\cite{Kneipp2006}, where ${\bf E}_{\rm loc}$ is the local electric-field  at the hotspot and ${\bf E}_0$ is the incident $\textbf{E}$-field  in free space. Impressive SERS enhancement factors as high as $10^{14}$ have been experimentally reported \cite{Kneipp2006}, which was observed in composites of metal nanoparticles (MNPs) that include dimers and small aggregates formed by MNPs \cite{Xu2000}, and fractal types of nanostructures \cite{Li2003}. The metal nanoparticles boost the electric-field by exploiting localized surface plasmon (LSP) resonances, that increase the LDOS and the effective excitation field strength. In the domain of single-molecule SERS \cite{Wang2013,Lee2013}, the sensitivity requirements are especially high, and so there is an incentive to further increase the Raman enhancement factors achievable by hotspots generated in plasmonic nanostructures. Other significant challenges include the generation of reproducible and stable hotspots and the accurate placement of a single molecule within/near an EM hotspot \cite{Wang2013,Lee2013}, which can also be aided, e.g., by novel nanostructures that can produce higher Raman enhancement without only resorting to decreasing the gap sizes between MNPs to increase $\textbf{E}$-field intensities.

In recent years, anisotropic optical materials  have been investigated extensively \cite{Poddubny2013,Ferrari2015,Shekhar2014}, especially those described using a diagonalized tensor with elements $\varepsilon_x$, $\varepsilon_y$, and $\varepsilon_z$ in which at least one of the elements exhibits a negative real part (and is thus metallic). Under certain conditions, these metal-dielectric hybrid materials, at least in bulk form, exhibit hyperboloid iso-frequency $\textbf{k}$-space surfaces, and are also referred to as indefinite media, with a dispersion relation given by
\begin{equation}
\label{eq:HMM_dispersion}
\frac{k_x^2 + k_y^2}{\varepsilon_\parallel} + \frac{k_z^2}{\varepsilon_\bot} = \frac{\omega^2}{c^2},
\end{equation}
where $\textbf{k}$ is the wavevector, $\omega$ is the angular frequency, and $c$ is the vacuum speed of light. Here, the axis of anisotropy corresponds to the $z$-axis. We refer to these general materials as hyperbolic materials (HMs). Equation \eqref{eq:HMM_dispersion} describes uniaxial HMs, which can be categorized into two types: Type I HMs, whose effective permittivity is characterized by $\varepsilon_x = \varepsilon_y = \varepsilon_\bot > 0$ and $\varepsilon_z = \varepsilon_\parallel < 0$; and Type II HMs, whose effective permittivity is characterized by $\varepsilon_x = \varepsilon_y = \varepsilon_\bot < 0$ and $\varepsilon_z = \varepsilon_\parallel > 0$. The dielectric tensor elements $\varepsilon_\parallel$ and $\varepsilon_\bot$ (parallel and perpendicular to the axis of anisotropy, respectively) are of opposite sign, corresponding to metallic or dielectric properties along different axes. The hyperboloid isofrequency surface enables access to very high values of $\textbf{k}$. The result of the high momentum mismatch between HM and free space EM fields implies strong confinement of light in the vicinity of the structure \cite{Yang2012}. In addition, a dipole emitter can couple to a large range of $\textbf{k}$-states at a given single frequency, which implies that the spontaneous emission (SE) rate of an embedded  emitter is highly enhanced in their vicinity \cite{Cortes2012}. One route to realize HMs is through artificial media, where a collection of  subwavelength sized elements of ordinary media can be organized to represent a single medium with hyperbolic dispersion properties \cite{Poddubny2013,Drachev2013,Ferrari2015}. For such artificial media, the unit cell size must be $d < \lambda_0/10$ in order for the effective medium approximation, i.e., by Maxwell-Garnett theory, to be valid \cite{Khaleque2015}. This is necessary to maintain the extent of the dispersion to high $\textbf{k}$-states, which in turn may lead to an ultra-high photonic LDOS (however, ultimately limited by material losses). There are also natural materials that display hyperbolic dispersion and are referred to as natural hyperbolic materials that have more recently been explored \cite{Narimanov2015,Korzeb2015}.

The enhanced LDOS of HMs may result in a large enhancement in the total SE rate of emitters located near the surface \cite{Jacob2012}, which   can also exceed that obtainable near pure metal surfaces. Most studies of HMs for SE enhancement have focused on emission enhancement of dipoles located inside or near the surface of the bulk material \cite{Gu2014,Noginov2010,Ni2011,Jacob2010,Newman2013}. Although the total SE decay rate is highly enhanced in the vicinity of HMs, the extraction of emitted light into free space is still a challenging problem due to optical quenching, and has many open questions \cite{Galfsky2015,Sreekanth2014,Axelrod2017}.

With regards to SERS, only a few special case studies have been investigated with artificial media \cite{Gaponenko2002,Zhang2015}. Gaponenko \cite{Gaponenko2002} summarized that the spontaneous Raman enhancement is attributed to the redistribution of the photonic LDOS over frequency, space, and directionality, being enhanced in specific regions of frequency, spatial location, and scattering direction. It was deduced that metal-dielectric based nanostructures would exhibit even higher relative variation of the LDOS, e.g., within a certain frequency range, compared to dielectric photonic crystals, and thus these artificial materials would exhibit higher spontaneous Raman enhancement. Zhang {\it et al.}~\cite{Zhang2015}
studied a hierarchical artificial material consisting of an array of 100 nm diameter nanoholes etched in gold that also consists of mesoscopic pores of 2 to 3 nm in size populating the entire bulk of the material, and fabricated and experimentally tested for its SERS enhancement properties. It was discovered that the achievable SERS enhancement can be superior to both the plain nanohole array and the mesoporous material without nanoholes, and the detection limit was below 1 $\times$ $10^{-13}$ M. While both of these works brought forward the use of artificial media with subwavelength metal-dielectric components in bulk form for the application to enhanced Raman spectroscopy, to our knowledge the use of this class of artificial media---an example of which is a HM---to construct nanoparticles for SERS, has not been investigated in the literature. Such HM based nanostructures have been investigated recently using planar stacks \cite{Yang2012}, nano-rod array \cite{Yao2011}, and concentric core-shell multilayers \cite{Wu2014}. The cited main advantage of these nano ``cavities'' is that the effective mode volume can be decreased substantially. Furthermore, HM nano-cavities exhibit relatively low quality ($Q$) factors, and thus offer a larger bandwidth with which to exploit the enhanced light-matter interaction. Despite these initial works, there does not seem to be a consensus on a design strategy that profits from the advantages offered by utilizing HMs for SERS, and there does not seem to be any physical insight into whether such materials would outperform metal nanostructures or not.

Recently, the use of HM nanostructures as a platform to develop single-photon sources was studied by Axelrod {\it et al.}\cite{Axelrod2017}, by investigating the achievable Purcell and single-photon $\beta$ factors (quantum efficiency) using a quasinormal mode approach. It was shown that despite the higher SE enhancement and LDOS achievable with HM based nanostructures, the radiative efficiency is much worse when compared to pure metals, which is captured by the very low single-photon $\beta$-factors and is mainly due to increased absorption and thus higher quenching in HMs \cite{Axelrod2017}. However, the scaling arguments and figures-of-merit for SERS are different than for a single photon source, i.e., the generated SERS is an intrinsically nonlinear phenomenon that requires an excitation field and the induced polarization exploits the total LDOS (as we will also show below, rather than just the radiative part).  Thus it is highly desirable to have a theory and analysis of SERS enhancement for HMs, firstly to understand the important figures-of-merit, and secondly to see if such structures can possibly outperform MNPs, even though they exhibit higher quenching. Such investigations also need to be complemented by considering the full propagation effects of light (i.e., from the Raman active molecule to a detector, usually in the far-field) as well as details of the nonlinear interplay between the molecule vibrations and the optical resonances. 

In this work, we present a detailed study on the use of LSP hotspots generated in HM nanostructures as a SERS platform. In Section \ref{Sec2}, the theoretical formalism of this work is described. We utilize the photon Green function to describe the optical reservoir properties of the nanoresonators (such as the LDOS), where the former is conveniently obtained using a semi-analytical quasinormal mode (QNM) theory \cite{Sauvan2013,Kristensen2014,Ge2014}. We then use this modal Green function to model the SERS spectra of molecules using a recently developed quantum optics approach for a general medium \cite{Dezfouli2017}. Section \ref{Sec3}  begins with a comparative study between our theoretical results and known experimental findings, which serves to emphasize the power of our formalism in accurately predicting SERS enhancement from first principles based on combining EM theory and quantum mechanics. Next, we investigate the SERS enhancement factors of several HM nanostructures and compare to their pure metal counterparts. A thorough account of the different contributing processes to the SERS enhancement is analysed and discussed. Although HM nanostructures can attain higher LDOS enhancements compared to their pure metal counterparts, the increased quenching leads to much lower SERS enhancement factors. Despite the fact that HMs cannot outperform traditional metal nanostructures for SERS when detected in the far-field, Section \ref{Sec4} will describe how the use of HMs can be beneficial when the technique of near-field probing of the SERS signal is utilized. In Sec.~\ref{Sec5}, we conclude with a summary and discussions of key findings and strategies for designing SERS devices that leverage the advantages offered by HMs while minimizing any negative effects. Finally, we present three Appendices that describe, respectively, the QNM Green function, the integration geometries used for the SERS calculations, and details of FDTD simulations and geometrical parameters of the various nanorod dimers that we investigate.

\section{Quantum Optics and Green Function Formalism TO\ Model the SERS Enhancement}
\label{Sec2}
Theoretical studies of SERS have been implemented using various approaches \cite{Kelley2008,Payton2013,Lombardi2014,Gu2007,Masiello2008,Masiello2010,Mueller2016,Delfan2012,Roelli2016,Schmidt2016}. On the one hand, some works describe SERS using semiclassical approaches (where the light is treated classically), and thus the complex interactions between molecule vibrations and the plasmonic field at the quantum level is not taken into account \cite{Gu2007,Delfan2012}, and only very specific optical structures and field regimes can be modeled. On the other hand, certain aspects of SERS can be considered using first-principles quantum mechanical description of the Raman active molecule through Density Functional Theory (DFT). This approach can also predict the electronic structure and thus the Raman modes {\it a priori}, which in turn can directly determine the Raman spectrum including effects of the local environment near the plasmonic metal nanoparticle \cite{Payton2013,Masiello2010}. However, DFT requires significant computational resources which may limit its applicability to real samples and the size of the problems tackled, as well as obscure the underlying physics of the SERS enhancement related to the optical modes and nonlinear field interactions. One of the most common limitations of the reports already published is that they largely do not delineate the contributions for enhancement of the excitation or scattering in the Raman scattering process. In this work, we employ a theoretical approach to calculate the SERS enhancement based on a quantum mechanical description of both the molecule vibrations and the plasmonic EM field \cite{Dezfouli2017}, which also allows us to consider realistic experimental configurations. Moreover, the technique is versatile and thus can be applied to different arbitrarily shaped metallic and HM plasmonic nanostructures, and provides an intuitive analytical understanding of the SERS process.

The general quantum optics theory to SERS is described elsewhere \cite{Dezfouli2017}, but here we just give the key results and concepts to be used in the present study. The theory uses an open system approach that allows one to derive analytical expressions for the detected Raman power spectral density (spectrum) of a molecule coupled to an arbitrary plasmonic nanostructure; moreover, the solution for both Stokes and anti-Stokes scattering contributions is given, which accounts for the spatial location of the detector and molecule with respect to the optical medium (e.g., the HM nanoresonator). In this formalism, both the molecule's vibrations and the EM fields are quantized, and their quantum mechanical interactions are taken into account. Furthermore, the full photon Green function is utilized in the theory, which rigorously captures the full frequency response of the plasmonic enhancement including quenching effects. As such, no modification to the theory is required for different nanostructures, but  rather the task is to obtain a suitable Green function for a given system of interest.

\begin{figure*}
        \centering
        \includegraphics[width=\textwidth]{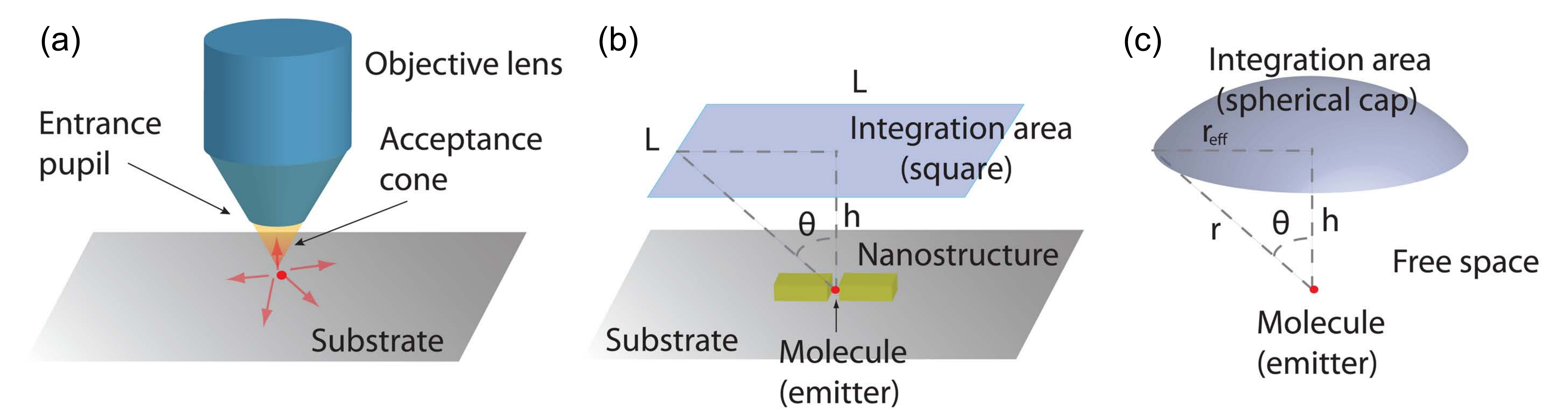}
        \caption{(a) Schematic of laser excitation and Raman scattered light collection using an objective lens. (b-c) Representation of the integration geometry for obtaining Raman scattered power when the molecule of interest is in the vicinity of a plasmonic nanostructure modeled using Lumerical FDTD, and when in free space treated analytically, respectively.}
        \label{fig:integration_geom}
\end{figure*}

A general definition of SERS enhancement factor for any configuration is extremely challenging if not impossible. However, a single molecule enhancement factor (SMEF) in a given SERS experiment can be defined as
\begin{equation}
\label{eq:SMEF}
\textnormal{SMEF} = \frac{P_{\textnormal{SERS}}^{\textnormal{SM}}(\omega_{\textnormal{st/as}})}{P_{\textnormal{RS}}^{\textnormal{SM}}(\omega_{\textnormal{st/as}})},
\end{equation}
where $P_{\textnormal{SERS}}^{\textnormal{SM}}(\omega_{\textnormal{st/as}})$ is the Raman scattered power [W] of a specific Raman mode (Stokes or anti-Stokes) of a given single molecule due to SERS enhancement  in the vicinity of the plasmonic nanostructure, and $P_{\textnormal{RS}}^{\textnormal{SM}}(\omega_{\textnormal{st/as}})$ is the Raman scattered power of the same molecule in free space \cite{Ru2007}. In an actual experiment, the Raman scattered light would be collected over a finite area corresponding to the size of the entrance pupil of the objective lens that is used to collect light into the Raman spectroscopy system, and thus only a portion of the total Raman power that is scattered at angles lying within the acceptance cone would be collected (Fig.\,\ref{fig:integration_geom}a). Furthermore, in a typical Raman setup, a spectrometer is used to carry out a frequency-dependent measurement in which the Raman scattered power is collected and separated into frequency bins, and thus the measured quantity would be the power spectral density [Ws] as a function of frequency. In order to obtain the scattered power of each Raman mode, the power spectral density must be integrated over the small range of frequencies over which the Raman scattering line covers. It is also important to note that $P_{\textnormal{SERS}}^{\textnormal{SM}}$ depends on several factors such as the location and orientation of the molecule with respect to the nanostructure geometry of the SERS substrate, and the direction and polarization of the excitation light with respect to the SERS substrate. Our calculation of $P_{\textnormal{SERS}}^{\textnormal{SM}}$  will be for the case yielding maximum detected Raman scattered signal; namely when the molecule is located at the central point of a particular hotspot generated near a metal nanostructure, and the orientation of the molecule is such that the direction of the dominant Raman tensor element is parallel to the direction of the dominant electric field at that point. A molecule position closer to the nanostructure surface, i.e., only a few nanometers away, may increase the total Raman scattering, but the increased quenching would reduce the detected Raman signal. Also, the polarization of the excitation light will be assumed parallel to the direction of the dominant Raman tensor element. Similarly, the Raman scattered power in free space $P_{\textnormal{RS}}^{\textnormal{SM}}$ is maximized when the polarization of the excitation field is parallel to the direction of the dominant Raman tensor element, and this is the quantity used for the SERS enhancement calculations below. 

The Raman scattering spectrum, scattered power of each Stokes/anti-Stokes resonance, and in turn the Raman enhancement of a molecule located in the vicinity of a plasmonic nanostructure can be calculated based on  the formalism highlighted above. The first step in the theory is to obtain the photon Green function (or reservoir function) $\textbf{G}(\textbf{r},\mathbf{r}^\prime;\omega)$ of the plasmonic nanostructure within its environment, which can be calculated based on the dominant resonant mode (or modes) that are determined here by a full-wave finite-difference time-domain (FDTD) method \cite{Lumerical}. Subsequently, the classical Green function is used to obtain the Raman spectrum that is derived using a quantum optics approach. Below we explain in more details how these quantities are computed and used.

\subsection{Classical Green function using QNM theory}

The Green function $\textbf{G}(\textbf{r},\mathbf{r}^\prime;\omega)$ is the solution to the classical Helmholtz equation with a point dipole source,
\begin{equation}
\label{eq:Helmholtz_eq}
\nabla \times \nabla \times \mathbf{G}(\textbf{r},\mathbf{r}^\prime;\omega) - \frac{\omega^2}{c^2}\varepsilon(\mathbf{r},\omega)\mathbf{G}(\textbf{r},\mathbf{r}^\prime;\omega) = \frac{\omega^2}{c^2}\mathbf{I}\delta(\mathbf{r} - \mathbf{r}^\prime), 
\end{equation}
subjected to, in general, open boundary conditions, where the relative permeability $\mu(\textbf{r},\omega) = 1$ as the materials considered here are non-magnetic. For a 3D metallic or HM nanostructure, different localized plasmonic modes can be excited at the corresponding resonant wavelengths that are determined by the permittivities and geometry of the structure,  and the polarization of the excitation light. Typically, only one QNM (LSP resonance) contributes dominantly to the scattering behavior of the nanostructure, which can be used to achieve plasmonic mode enhancement of the SERS signal. These localized plasmonic resonances are formally quasinormal modes (QNMs) \cite{Kristensen2014}, which are solutions to a non-Hermitian Maxwell's equation subjected to open boundary conditions \cite{Leung1994,Lee1999}. These QNMs have complex eigenfrequencies $\tilde{\omega}_\mu = \omega_\mu - i\gamma_\mu$, where the imaginary part quantifies the energy loss through  $Q_{\mu} = \omega_\mu/(2\gamma_\mu)$. As shown elsewhere, the system Green function can be accurately represented through an expansion of these QNMs, $\tilde{\mathbf{f}}_{\mu}(\mathbf{r})$, using \cite{Ge2014}
\begin{equation}
\label{eq:Green_function}
\mathbf{G}(\textbf{r},\mathbf{r}^\prime;\omega) = \sum_\mu\frac{\omega^2}{2\tilde{\omega}_\mu(\tilde{\omega}_\mu - \omega)}\,\tilde{\bf f}_\mu(\mathbf{r})\tilde{\bf f}_\mu(\mathbf{r}^\prime), 
\end{equation}
where the QNMs are appropriately normalized. Using a single QNM expansion with $\tilde{\bf f}_c$, then one simply has
\begin{equation}
\label{eq:Green_function_single}
\mathbf{G}(\textbf{r},\mathbf{r}^\prime;\omega) = \frac{\omega^2}{2\tilde{\omega}_c(\tilde{\omega}_c - \omega)}\,\tilde{\bf f}_c(\mathbf{r})\tilde{\bf f}_c(\mathbf{r}^\prime).
\end{equation}

In this paper, we use the QNM normalization scheme \cite{Leung1994,Lee1999}
\begin{align}
& \langle\langle\tilde{\bf f}_\mu|\tilde{\mathbf{f}}_\mu\rangle\rangle = \lim_{V \to \infty}\left\{\int_V\tilde{\bf f}_\mu(\mathbf{r})\cdot\left[\frac{\partial(\omega\varepsilon(\mathbf{r},\omega))}{\partial\omega}\right]_{\tilde{\omega}_\mu}
\cdot\tilde{\bf f}_\mu(\mathbf{r})d\mathbf{r} \right. \nonumber \\
& \quad\quad + \left.\frac{ic}{2\tilde{\omega}_\mu}\int_{\partial V}\tilde{\bf f}_\mu(\mathbf{r})\cdot\sqrt{\varepsilon(\mathbf{r},\omega_\mu)}\cdot\tilde{\bf f}_\mu(\mathbf{r})d\mathbf{r}\right\} = 1,\label{eq:norm_factor}
\end{align}
which, in addition to a  volume integration similar to what one sees in normal mode theories, there is also a surface integration on the boundaries. To obtain the QNMs of the desired system, we employ   Lumerical  FDTD  \cite{Lumerical} and follow the QNM recipe given in Ref.\,\onlinecite{Ge2015}, being careful of how to perform the integration \cite{Kristensen2015}. A comparison of the performance between our QNM theory and fully numerical solutions of Maxwell's equations is given in Appendix \ref{appendix:Green_function}. We note that although the QNM approach is not needed for the designs below, it helps to clarify the underlying physics in terms of the domonant LSP modes. Moreover, using such a modal approach, one can easily compute the SERS from molecules located at any spatial position, without having to do any additional point dipole simulations in Maxwell's equations.

\subsection{Quantum optomechanical formalism of SERS}

Given a known Raman tensor and its dominant element $R_{nn}$ oriented along the unit vector $\textbf{n}$, which is an intrinsic property of the Raman mode, and is directly related to the Raman cross-section, the Raman scattering spectrum can be calculated for a molecule of known orientation with respect to the SERS substrate geometry and the polarization of the excitation field. Consider a SERS experiment where a continuous-wave (CW) laser with electric field amplitude $\mathbf{E}_0$ and frequency $\omega_L$, excites the molecule with an active oscillation mode at frequency $\omega_m$ placed nearby the plasmonic nanostructure. The Raman scattering spectrum from a molecule at spatial location $\textbf{r}_m$ detected at a certain location $\textbf{r}_D$ is derived to be \cite{Dezfouli2017} 
\begin{equation}
\label{eq:Emission_spect}
S(\mathbf{r}_D,\omega) = A(\omega_L,\mathbf{E}_0)S_0(\omega)|\mathbf{G}(\mathbf{r}_D,\mathbf{r}_m;\omega) \cdot \mathbf{n}|^2,    
\end{equation} 
where, in general, three different processes are involved. First, the prefactor $A(\omega_L,\mathbf{E}_0)$ quantifies the plasmonic enhancement of the induced Raman dipole through $A(\omega_L,\mathbf{E}_0) = \hbar R_{nn}^2 |\eta|^2 |\mathbf{n} \cdot \mathbf{E}_0|^2 / (2\omega_m\varepsilon_0^2)$, where $\eta$ is the field enhancement factor at the molecule location $\mathbf{r}_m$ due to the presence of the scattering geometry. The enhanced field is  the effective excitation field seen by the molecule (self-consistent scattering solution) that can be calculated using the  Green function of Eq.\,\eqref{eq:Green_function}:
\begin{equation}
\label{eq:field_enhancement}
\eta = 1 + \frac{\int_V\mathbf{n}\cdot[\varepsilon(\mathbf{r},\omega_L) - \varepsilon_B]\cdot\mathbf{G}(\mathbf{r}_m,\mathbf{r};\omega_L)\cdot\mathbf{E}_0(\mathbf{r})d\mathbf{r}}{\mathbf{n}
\cdot\mathbf{E}_0(\mathbf{r}_m)},
\end{equation}
where the spatial integration runs over the finite-size scattering geometry (i.e., the metal) and $\varepsilon_B$ is the background medium dielectric function. Here, we also assume that $\textbf{E}_0(\mathbf{r})$ is  constant for all spatial positions $\textbf{r}$, which is an accurate approximation since the diameter of the laser beam waist is typically much larger than the size of the plasmonic nanostructure. The second term in Eq.\,\eqref{eq:Emission_spect}, $S_0(\omega)$, represents the $\textit{emitted}$ spectral function for the Raman scattered photons that is calculated using quantum optical modeling of the molecule-field dynamics where the plasmonic system is treated as a photonic reservoir and the molecule is a quantized harmonic oscillator; physically, it describes the plasmonic enhancement of the intrinsic Raman scattering rate of a molecule, which is given by $S_0(\omega) = S_0^{\textnormal{st}}(\omega) + S_0^{\textnormal{as}}(\omega)$, where the Stokes and anti-Stokes contributions are defined analytically as follows \cite{Dezfouli2017}:
\begin{align}
& S_0^{\textnormal{st}}(\omega) = \label{eq:S0_st} \\
& {\rm Re}\left\{\frac{i[\gamma_m(\bar{n}^{\textnormal{th}} + 1) + J_{\textnormal{ph}}(\omega_L + \omega_m)]}{[\omega - (\omega_L - \omega_m) + i(\gamma_m +\Delta J_{\textnormal{ph}})](\gamma_m + \Delta J_{\textnormal{ph}})}\right\}, \nonumber \\
& \nonumber \\
& S_0^{\textnormal{as}}(\omega) = \label{eq:S0_as} \\
& {\rm Re}\left\{\frac{i[\gamma_m\bar{n}^{\textnormal{th}} + J_{\textnormal{ph}}(\omega_L - \omega_m)]}{[\omega - (\omega_L + \omega_m) + i(\gamma_m +\Delta J_{\textnormal{ph}})](\gamma_m + \Delta J_{\textnormal{ph}})}\right\} \nonumber.
\end{align}
Here, $\gamma_m$ is the decay rate of the Raman vibrational mode, and $\bar{n}^{\textnormal{th}}=\left(e^{\hbar\omega_m/k_B T} - 1\right)^{-1}$ is the thermal population of the Raman vibrational mode (at some temperature $T$). The function $J_{\textnormal{ph}}$ describes the plasmonic-induced Raman scattering rate beyond that achievable through thermal population, which depends on the projected LDOS through an effective bath spectral function,
\begin{equation}
\label{eq:Jph}
J_{\textnormal{ph}}(\omega) = \frac{R_{nn}^2|\eta|^2 |\mathbf{n} \cdot \mathbf{E}_0|^2}{2\varepsilon_0\omega_m}\textnormal{Im}\{\mathbf{G}_{nn}(\mathbf{r}_m,\mathbf{r}_m;\omega)\},    
\end{equation}
and we have defined in Eqs.\,(\ref{eq:S0_st}) and (\ref{eq:S0_as}): $\Delta J_{\textnormal{ph}} = J_{\textnormal{ph}}(\omega_L + \omega_m) - J_{\textnormal{ph}}(\omega_L - \omega_m)$. The third and final term in Eq.\,\eqref{eq:Emission_spect}, $|\mathbf{G}(\mathbf{r}_D,\mathbf{r}_m;\omega) \cdot \mathbf{n}|^2,$  includes the enhancement of the scattered field and the appropriate propagation effects from the molecule location to the detector, again using the  Green function. 

To come closer to actual experiments, finally we note that the SERS power at the detector is obtained by first integrating the SERS spectrum of Eq.\,\eqref{eq:Emission_spect} over the detector area,  followed by an integration over the spectral range $\Delta\omega$ that the Raman scattering line covers, so we define\begin{equation}
\label{eq:SERS_intensity_calc}
P_{\textnormal{SERS}}^{\textnormal{SM}}(\omega_{\textnormal{st/as}}) = \frac{\varepsilon_0 c n}{2} \int_{\Delta\omega}\int_{\rm det}S(\mathbf{r}_D,\omega)dA\,  d\omega,
\end{equation} 
where $n=\sqrt{\varepsilon_B}$ is the refractive index of the medium. The prefactors in Eq.\,(\ref{eq:SERS_intensity_calc}) are necessary to give $P_{\textnormal{SERS}}^{\textnormal{SM}}$ the correct standard units of power, namely [W].

In order to obtain the single molecule enhancement factor defined in Eq.\,\eqref{eq:SMEF}, we first consider the molecule scattering  when in the vicinity of the plasmonic nanostructure, and then again in free space. Both of these scenarios can naturally be computed by the formalism discussed above, where the Green function of the plasmonic device and the free space Green function are used, respectively. The general Raman schemes we model are shown schematically  in Fig.\,\ref{fig:integration_geom}, where two different integration geometries are considered. More details on the setup of the integration geometries for the calculations of both $P_{\textnormal{SERS}}^{\textnormal{SM}}$ and $P_{\textnormal{RS}}^{\textnormal{SM}}$ are given in Appendix \ref{appendix:Integration_geom}.

In what follows we shall explore the value and efficacy of this model by predicting the SERS enhancement of gold nanorod dimers, and then elucidating how these compare to SERS enhancement using HMs.

\section{SERS enhancement in Au nanorod dimers and hyperbolic material dimers}
\label{Sec3}
A MNP such as a metallic nanorod with characteristic dimensions on the order of tens to hundreds of nanometers would exhibit a LSP resonance at a wavelength $\lambda_\textnormal{LSP}$ within the visible to near-infrared wavelengths. A metal dimer is typically formed when two separate MNPs are placed close to each other with a separation gap of only tens of nanometers.  The enhanced $\textbf{E}$-field in the gap of a dimer and thus increased LDOS near the surface of a MNP are both crucial for SERS enhancement. Here we explore the SERS enhancement of both pure gold (Au) and HM nanorod dimers, where the Type-II HM based on an alternating multilayer stack of Au and $\textnormal{Si}_3\textnormal{N}_4$ is utilized. The wavelength-dependent complex refractive indices of Au and $\textnormal{Si}_3\textnormal{N}_4$ are taken from  Johnson and Christy \cite{Johnson1972} and Philipp \cite{Philipp1973}, respectively. It should be noted that we do not use an effective medium approximation in this work, but rather the true spatial dependence of the dielectric function is incorporated in all of our investigations below.

\subsection{Prediction of known experimental SERS enhancement in metal nanoresonators}

\begin{figure}
        \centering
        \includegraphics[width=\columnwidth]{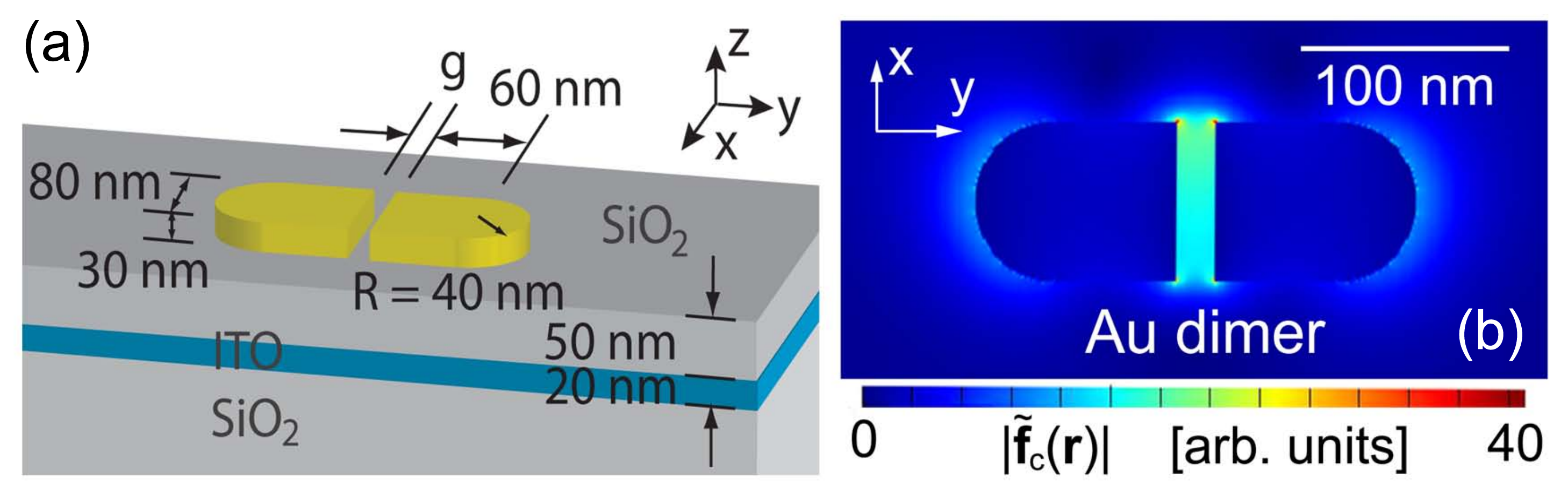}
        \caption{(a) Schematic of Au nano-dimer, as in Ref.\,\onlinecite{Zhu2011}, that we use for theoretical comparison given in Table \ref{table:enhancements_comp_exp}. (b) The corresponding field amplitude $|\tilde{\textbf{f}}_c(\textbf{r})|$ of the localized surface plasmon at the resonant $\lambda_\textnormal{LSP}$ at $z$-plane intersecting center of dimer gap, calculated and normalized using QNM theory.}
        \label{fig:dimer_geom}
\end{figure}

In this subsection, we examine the use of the  approach described in Section \ref{Sec2} to predict the SERS performance in typical experimental configurations. The Au dimer structure in Ref.\,\onlinecite{Zhu2011} is used as a representative example to compare our theoretical to the experimental results obtained for the SERS enhancement. The experimentally characterized Au dimers have dimensions that are approximated by those schematically shown in Fig.\,\ref{fig:dimer_geom}a. The Au dimer sits on top of a layered substrate consisting of 50 nm $\textnormal{SiO}_2$/20 nm ITO/$\textnormal{SiO}_2$. The gap size $\textit{g}$ is varied, and in Ref.\,\onlinecite{Zhu2011} gap sizes ranging from approximately 3 nm to 20 nm have been studied, while here gap sizes of 10 nm and 20 nm are investigated for comparison to the experimental results. In Ref.\,\onlinecite{Zhu2011}, SERS measurements are carried out with a confocal Raman microscope, in which the Raman signal is collected by a 20$\times$ objective lens (NA = 0.4) and directed into a spectrometer with a CCD array. Prior to measurements, the sample containing Au dimers is immersed in a 3 mM solution of benzenethiol to form a self-assembled monolayer, and thus covering all sides of each Au dimer. 

In order to determine the SERS enhancement factor (EF), reference measurements are made on a sample cell containing pure benzenethiol. The experimental SERS EF is determined from \cite{Zhu2011}
\begin{equation}
\label{eq:SERS_EF_exp}
\textnormal{EF} = 
\frac{P_{\textnormal{SERS}}/N_{\textnormal{SERS}}}{P_{\textnormal{RS}}/N_{\textnormal{RS}}},
\end{equation}
where $P_{\textnormal{SERS}}$ and $P_{\textnormal{RS}}$ are the measured Raman signal powers of a specific Raman scattering line for the SERS substrate and reference sample, respectively. As in the calculation of $P_{\textnormal{SERS}}^{\textnormal{SM}}$ in Eq.\,\eqref{eq:SERS_intensity_calc}, each of $P_{\textnormal{SERS}}$ and $P_{\textnormal{RS}}$ is also computed by integrating the power spectral density (obtained from the measured Raman spectrum) over the range of frequencies that the Raman line of interest covers. The resulting Raman signal power is then processed by removing the background signal, and also normalized to the incident laser power and CCD integration time. The $N_{\textnormal{SERS}}$ and $N_{\textnormal{RS}}$ are the number of probed benzenethiol molecules in the excitation laser spot for the SERS substrate and reference sample, respectively. To determine $N_{\textnormal{SERS}}$, first the surface density of benzenethiol is assumed based on the chemistry of the self-assembly process, and then the surface area of each Au dimer including all top surfaces and sidewalls, as well as the number of Au dimers within the laser spot area are considered. Finally, $N_{\textnormal{RS}}$ is obtained by considering the volume of the excitation laser spot within the reference sample cell containing benzenethiol and also the number density of molecules in the sample. More details of the experimental SERS EF calculation are given in the supporting information of Ref.\,\onlinecite{Zhu2011}.

\begin{table}
                
               \begin{tabular}{ | c | c | c | c |}
                \hline
                Au dimer gap & $\lambda_\textnormal{LSP}$ [nm] & SMEF (theory) & SERS EF (exp.) \\ \hline
                10 nm & 763.7 & 9.42 $\times 10^7$ & $\sim 4.0 \times 10^6$ \\ \hline
                20 nm & 728.5 & 3.49 $\times 10^7$ & $\sim 1.5 \times 10^6$ \\ \hline
               \end{tabular}
               \caption{Comparison between our theoretical SMEF versus the experimentally reported SERS EF in Ref.\,\onlinecite{Zhu2011} for two different nano-dimer gap sizes, as indicated in Fig.\,\ref{fig:dimer_geom}. The corresponding values of $\lambda_\textnormal{L}$ are 733 nm and 700 nm for the gap sizes of 10 nm and 20 nm, respectively. The values of $\lambda_\textnormal{st}$ are 795.5 nm and 756.8 nm for the gap sizes of 10 nm and 20 nm, respectively. While it is difficult to compare absolute numbers (e.g., because of detection geometries), the general trend is seen to be very good and substantially closer to the experimental values and trends compared to other SERS theories (see text).  }
               \label{table:enhancements_comp_exp}
                
\end{table}

The theoretical calculation of SMEF reflects the experimental conditions in that $\lambda_\textnormal{L}$ is chosen based on $\lambda_\textnormal{LSP}$ obtained from Lumerical FDTD simulations \cite{Lumerical}. The values of $\lambda_\textnormal{LSP}$ for the Raman line with shift $\nu = 1072~ \textnormal{cm}^{-1}$ ($\omega_m$ = 2.02 $\times 10^{14}$ rad/s) are given in Table \ref{table:enhancements_comp_exp} for the Au nano-dimers with $\textit{g}$ = 10 nm and 20 nm. The corresponding values of $\lambda_\textnormal{L}$ are thus 733 nm and 700 nm for the Au nano-dimers with $\textit{g}$ = 10 nm and 20 nm, respectively. The SMEF calculated theoretically and the experimental SERS EF for these example Au nano-dimer structures are also shown in Table \ref{table:enhancements_comp_exp}. As can be noted, the theoretical value of the SMEF is $\sim$23$\times$ higher compared to the experimental SERS EF, for both the Au nano-dimers with gaps of 10 nm and 20 nm. Given the inherent repeatability challenges associated with SERS measurement, the assumptions made in the calculations presented here about the optimal location of the Raman molecule, and the lack of any fitting parameters in the model, this difference is remarkably small when compared to what was reported previously in the literature. More remarkable is the versatility of this approach, which is obtained from first principles and involves self-consistent calculations without resorting to any artificial fitting parameters. The  apparent over-estimation of the theoretical SMEF compared to the experimental SERS EF is also expected, and is attributed to the fact that SMEF represents the SERS enhancement for a single molecule located at the ideal location, which is the center of the Au nano-dimer gap where the hotspot intensity is the highest, and thus it would be significantly higher than the average SERS enhancement experienced by the molecules that cover the surface of the nano-dimer, which is the quantity of the experimental EF. Furthermore, in the theoretical calculation of SMEF, the orientation of the molecule is assumed such that the dominant Raman tensor element is aligned parallel to the nano-dimer axis, which is also the dominant $\textbf{E}$-field polarization of the LSP resonance, thus resulting in maximized SERS enhancement (the theory value is thus an upper bound). However, in an actual experiment, the orientations of molecules near the surface of the nano-dimer are random, which means the measured SERS EF is an average of those obtained for different molecule orientations. What must be emphasized here is the capability of our technique for making direct and reliable comparison between different devices and designs. Note that the ratio between SERS enhancement factors of the two nano-dimers shown in Table \ref{table:enhancements_comp_exp} is roughly the same both by using our theoretical investigation presented and from the experimental values reported in Ref.\,\onlinecite{Zhu2011}, namely 2.7 and 2.67 for the former and latter cases, respectively. It is also important to point out that in most previous work on theoretical accounts of SERS enhancement, the calculated values are typically lower than the experimentally determined SERS EF by 1 to 2-orders of magnitude, which have been qualitatively attributed to chemical enhancement that has not been considered \cite{Zhu2011,Gu2007,Zhu2014,Wustholz2010,Talaga2015,Lim2011}. The conventional approach to estimating the SERS enhancement is through the quantity $(|\textbf{E}|/|\textbf{E}_0|)^4$\,\,\cite{Kneipp2006} at $\lambda_\textnormal{LSP}$ of the plasmonic resonance, which we have calculated to be 5.89 $\times$ $10^5$ and 8.55 $\times$ $10^4$, taken at the center of the nano-dimer gap of 10 nm and 20 nm, respectively. It is seen that the $(|\textbf{E}|/|\textbf{E}_0|)^4$ enhancement is approximately 1-order of magnitude lower than the experimental EF for both cases of the nano-dimer gaps of 10 nm and 20 nm. Although the magnitude of error using $(|\textbf{E}|/|\textbf{E}_0|)^4$ rule in predicting the experimental EF is similar to that based on our approach, it must be emphasized that an under-estimation is problematic because it requires qualitative explanation using chemical enhancement. Clearly, the SERS enhancement calculated in this work---using a simple analytical quantum optics approach and EM theory---is sufficient to predict the experimentally determined enhancement without the need to invoke that chemical enhancement is also involved; this approach is different from conventional wisdom that is simply based on the assumed $\textbf{E}$-field enhancement.

\subsection{Performance of hyperbolic materials for SERS}

One of the central advantages of HMs reported in the literature is the enhanced LDOS that serves to improve, amongst other applications, the total SE rates of molecules that sit near the HM \cite{Cortes2012}. Previous work has mostly focused on investigating LDOS enhancement inside or near the surface of bulk HMs \cite{Gu2014,Noginov2010,Ni2011,Jacob2010,Newman2013}, but here we rigorously study the effect for HM based nanostructures that exhibit the LSP resonance. As such, we will use our approach to study the SERS performance of HM nanorod dimers in order to delineate in detail whether HMs provide any advantages over regular MNPs with comparable geometries. 

\begin{figure}
        \centering
        \includegraphics[width=\columnwidth]{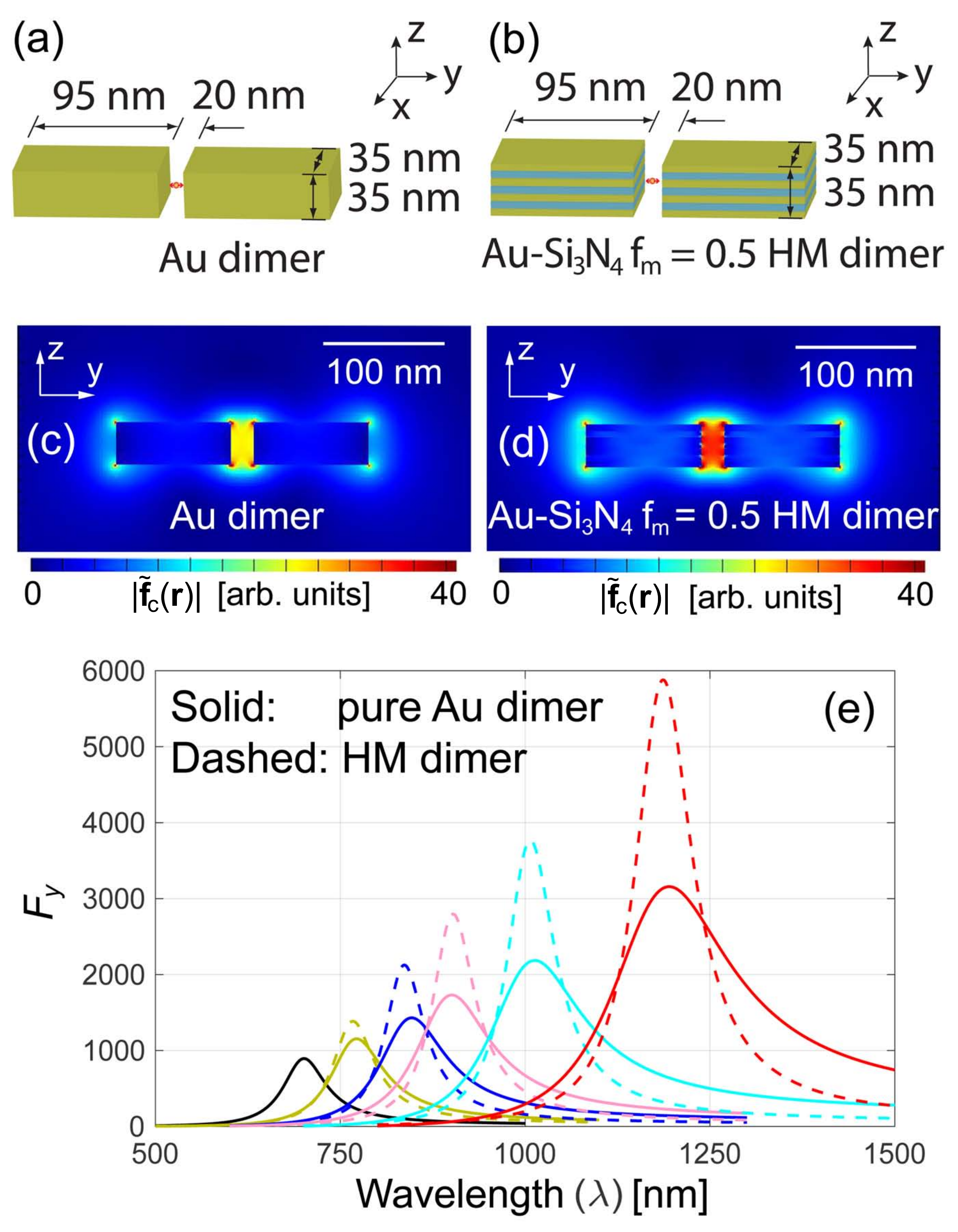}
        \caption{Schematics of (a) Au nanorod dimer, and (b) $\textnormal{Au}$-$\textnormal{Si}_3\textnormal{N}_4$ $f_m$ = 0.5 HM nanorod dimer. Electric-field amplitude $|\tilde{\bf f}_c(\textbf{r})|$ of localized surface plasmon at the resonant $\lambda_\textnormal{LSP}$ at x-plane intersecting center of dimer gap for (c) Au nanorod dimer (95 nm nanorod length), and (d) $\textnormal{Au}$-$\textnormal{Si}_3\textnormal{N}_4$ $f_m$ = 0.5 HM nanorod dimer (95 nm nanorod length). (e) Purcell factor at gap center of Au and HM nanorod dimers. Solid lines show Au nanorod dimers with nanorod lengths of 95 nm (black), 115 nm (green), 135 nm (blue), 150 nm (gray), 180 nm (cyan), and 230 nm (red). Dashed lines show HM nanorod dimers (95 nm nanorod length) with $f_m$ = 0.75 (green), 0.5 (blue), 0.4 (gray), 0.3 (cyan), and 0.2 (red).}
        \label{fig:Purcell_factor}
\end{figure}

Based on QNM theory, the Green function of several HM nanorod dimers are obtained, from which the LDOS is calculated. The results are then compared to that of the reference pure-Au nanorod dimer structure. The HM is characterized by the volume fraction of metal within the material, or the metal filling fraction $f_m$, which determines the anisotropic complex permittivity. When $f_m$ is decreased while maintaining the same overall nanorod dimer dimensions, the LSP wavelength $\lambda_\textnormal{LSP}$ is red-shifted. To ensure a fair comparison of the LDOS between pure Au and HM nanorod dimers, additional pure Au dimers with increased lengths are designed to have $\lambda_\textnormal{LSP}$ that correspond (or very close to) the resonant wavelengths of the HM nanorod dimers. A summary of the dimensions of the reference Au nanorod dimer and the HM nanorod dimer with $f_m$ = 0.5 is schematically shown in Fig.\,\ref{fig:Purcell_factor}. The details of the dimensions of individual Au and $\textnormal{Si}_3\textnormal{N}_4$ layers within each HM nanorod, and the nanorod lengths of the different Au nanorod dimers with corresponding $\lambda_\textnormal{LSP}$ are described in Appendix \ref{appendix:sim_setup}. The potential strategies for fabricating these stratified HM nanorod dimers are also given in Appendix \ref{appendix:sim_setup}. Note that while the nanostructures used for SERS in experiments typically sit on top of substrates for practical fabrication related reasons, the nanorod dimer structures we investigate in this work are in free space as this configuration does not affect the basic physics of the SERS enhancement, and it is computationally less intensive to simulate.

For each nanorod dimer, Lumerical FDTD is employed to perform a full-dipole simulation to determine the LSP wavelength $\lambda_{\textnormal{LSP}}$ as well as its quality factor through Lorentzian fitting of the frequency response. The QNM is then obtained by performing another simulation in which a $\textit{y}$-polarized plane wave excitation is incident in the $\textit{z}$-direction, and using a 3-dimensional field monitor to capture the corresponding QNM. The QNM approach is verified against full dipole calculations by investigating the SE enhancement rate of a quantum emitter at the dimer gap center in each case, where good agreement (within a few percent) was achieved (see Appendix \ref{appendix:Green_function}), and the difference is attributed to numerical imperfections and possible contributions from higher order QNMs. Further details of the FDTD simulations setup are given in Appendix \ref{appendix:sim_setup}.

Quite generally, for any structure, the Green function (obtained here using QNM theory) can be used to obtain the SE enhancement factor, $F$, for a dipole emitter located at $\mathbf{r}_0$, polarized along $\mathbf{n}$, and placed in the background medium with refractive index of $n_B$, through
\begin{equation}
\label{eq:Purcell_factor}
F\left(\mathbf{r}_0;\omega\right) = \frac{6\pi c^3}{\omega^3 n_B}\,\textnormal{Im}\{\mathbf{n}\cdot\mathbf{G}(\mathbf{r}_0,\mathbf{r}_0;\omega)\cdot\mathbf{n}\}.
\end{equation}
This generalized Purcell factor (enhanced emission factor) is obtained by 
\begin{equation}
\label{eq:Purcell_factor_2}
F_\textbf{n}\left(\mathbf{r}_0;\omega\right) = \frac{\textnormal{Im}\{\mathbf{n}\cdot\mathbf{G}(\mathbf{r}_0,\mathbf{r}_0;\omega)\cdot\mathbf{n}\}}{\textnormal{Im}\{\mathbf{n}\cdot\mathbf{G}_0(\mathbf{r}_0,\mathbf{r}_0;\omega)\cdot\mathbf{n}\}},
\end{equation}
where $\textbf{G}_0$ is the free space Green function. The Purcell factor when the dipole is oriented in the $y$-direction $F_y$ is plotted in Fig.\,\ref{fig:Purcell_factor}e for all of the structures we study. Results for pure gold and HM dimers are plotted in solid and dashed lines, respectively, where different colors represent different metal filling fractions and sizes as explained in the figure caption. It is observed that the peak enhancement increases when the metal filling fraction of the HM dimer decreases, and also when increasing the length of the Au dimer, which both serve to red-shift the resonance peak. The peak $F_y$ at $\lambda_{\textnormal{LSP}}$ increases by over 6 times from approximately 893 for the Au nanorod dimer with $f_m = 1$ ($\lambda_{LSP}$ = 699.7 nm) to 5880 for the HM nanorod dimer with $f_m = 0.2$ ($\lambda_{LSP}$ = 1185.4 nm). Comparing the Au and HM nanorod dimers with $\lambda_{LSP}$ $\approx$ 1190 nm, the HM nanorod dimer with $f_m$ = 0.2 attains a peak $F_\textit{y}$ that is almost 2 times that of the Au nanorod dimer with a nanorod length of 230 nm. As such, in terms of total LDOS enhancement, HMs are clearly superior to pure MNPs. The absolute LDOS for all of the Au and HM nanorod dimers investigated here are also presented in Appendix \ref{appendix:Green_function}. 

\begin{figure*}
        \centering
        \includegraphics[width=\textwidth]{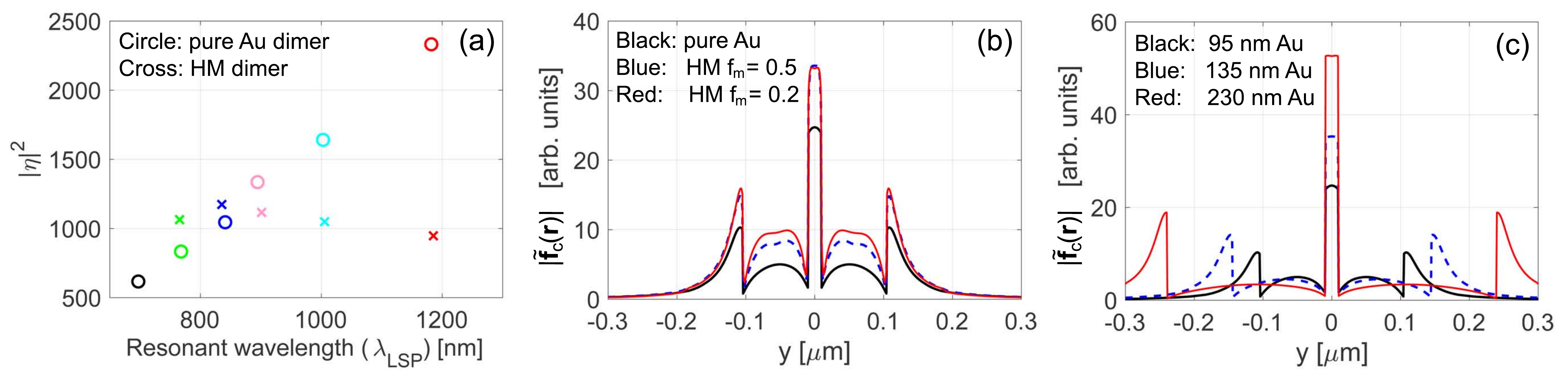}
        \caption{(a) Square of field enhancement factor at dimer gap center as a function of resonant wavelength $\lambda_\textnormal{LSP}$ for Au and HM nanorod dimers. Circles show Au nanorod dimers with nanorod lengths of 95 nm (black), 115 nm (green), 135 nm (blue), 150 nm (gray), 180 nm (cyan), and 230 nm (red). Crosses show HM nanorod dimers (95 nm nanorod length) with $f_m$ = 0.75 (green), 0.5 (blue), 0.4 (gray), 0.3 (cyan), and 0.2 (red). (b-c) Electric field amplitude $|\tilde{\bf f}_c(\textbf{r})|$ of the dominant QNM as a function of location along the $\textit{y}$-axis intersecting the center of the nanorod dimer gap, when comparing the different metal filling fractions for HMs and different nanorod lengths for pure Au, respectively.}
        \label{fig:field_enhancement}
\end{figure*}

Referring to Fig.\,\ref{fig:Purcell_factor}(b), it can be seen that the direction perpendicular to the HM's axis of anisotropy is also parallel to the nanorod dimer axis that is the dominant direction of polarization for the LSP ($y$-direction). Decreasing $f_m$ of the HM actually increases the wavelength at which the permittivity in the direction perpendicular to the HM's axis of anisotropy $\varepsilon_\bot$ (corresponding to the direction of dominant LSP polarization) equals zero, also referred to as the epsilon-at-zero wavelength. The epsilon-at-zero wavelength is directly related to the surface plasmon (SP) wavelength, and thus it is also red-shifted for decreasing $f_m$. This is partially what contributes to the red-shift of the HM nanorod dimer LSP resonance peak as $f_m$ is decreased [Fig.\,\ref{fig:Purcell_factor}(e)]. However, it must be noted that by decreasing the $f_m$ of the HM from 0.75 to 0.2, while the epsilon-at-zero wavelength for $\varepsilon_\bot$ red-shifts from $\sim$412 nm to $\sim$695 nm, the LSP resonance peak of the nanorod dimer red-shifts by a much larger extent from $\sim$699.7 nm to $\sim$1185.4 nm. As such, there is no simple relationship between the anisotropic permittivity of the HM as dictated by its $f_m$ and the LSP resonant wavelength of the HM based nanostructure, as other factors including the size and shape of the nanostructure together determine its optical properties.

As shown in Section \ref{Sec2}, the Raman scattering spectrum is intimately dependent on the total Green function of the photonic system under consideration. Indeed, the Green function comes into play in a much more complicated fashion in determining the SERS enhancement in comparison to the simply LDOS in the Purcell factor. One of the powerful attributes of the quantum optics approach we utilize here is that it can clearly separate the different contributions to the SERS enhancement. For example, the contribution by the enhancement of the excitation intensity is given by $|\eta|^2$, where $|\eta|$ represents the $\textbf{E}$-field enhancement factor. From studying Fig.\,\ref{fig:field_enhancement}a, it can be seen that for the Au nanorod dimers, $|\eta|^2$ increases linearly as a function of the resonant wavelength $\lambda_{\rm LSP}$, which is the same trend as for the Purcell factor. However, for the HM nanorod dimers, $|\eta|^2$ first increases as $f_m$ is decreased (i.e., an increase in $\lambda_\textnormal{LSP}$), but then decreases again upon further reducing $f_m$. The value of $|\eta|^2$ is maximized for the HM nanorod dimer with $f_m$ $\approx$ 0.5. This deviation in behavior compared to LDOS enhancement shows that additional physical effects are at play in the field enhancement process. It is also interesting to note that the field enhancement factor attained by utilizing HM nanorod dimers can be higher than with the Au nanorod dimer counterpart with identical $\lambda_\textnormal{LSP,}$ only for a certain range of $f_m$ approximately from 0.45 to approaching 1 (pure Au). For very low $f_m$ HMs, the field enhancement is much worse than by simply using Au nanorod dimers.

While the LDOS increases by using HM resonators instead of pure metal (i.e., by decreasing $f_m$), absorption within the nanorod dimer simultaneously increases due to higher $\textbf{E}$-field amplitude within the lossy nanorods, which is shown in Fig.\,\ref{fig:field_enhancement}b for several HM nanorod dimers. This increase in quenching results in a decrease in the $\textbf{E}$-field enhancement as $f_m$ of the HM nanorod dimer reduces to below $\approx$ 0.5. From Fig.\,\ref{fig:field_enhancement}c, it is also seen that by increasing the length of the individual nanorods of the Au nanorod dimer, the $\textbf{E}$-field amplitude within the nanorods does not change appreciably and thus quenching remains approximately constant, which is why $|\eta|^2$ as shown in Fig.\,\ref{fig:field_enhancement}a for the Au nanorod dimers still increases monotonically as a function of $\lambda_\textnormal{LSP}$. One also sees that the $\textbf{E}$-field amplitude of the QNM inside the HM nanorod dimer gap region increases as $f_m$ decreases from 1 to 0.5, but then decreases slightly as $f_m$ decreases from 0.5 to 0.2. However, increasing the Au nanorod length from 95 nm to 230 nm to produce the same red-shift in $\lambda_\textnormal{LSP}$ increases the $\textbf{E}$-field amplitude in the gap region by more than 2 times. Therefore, unlike the LDOS enhancement, the $\textbf{E}$-field enhancement process is negatively affected by quenching due to material absorption within the nanorods.

\begin{figure*}
        \centering
        \includegraphics[width=\textwidth]{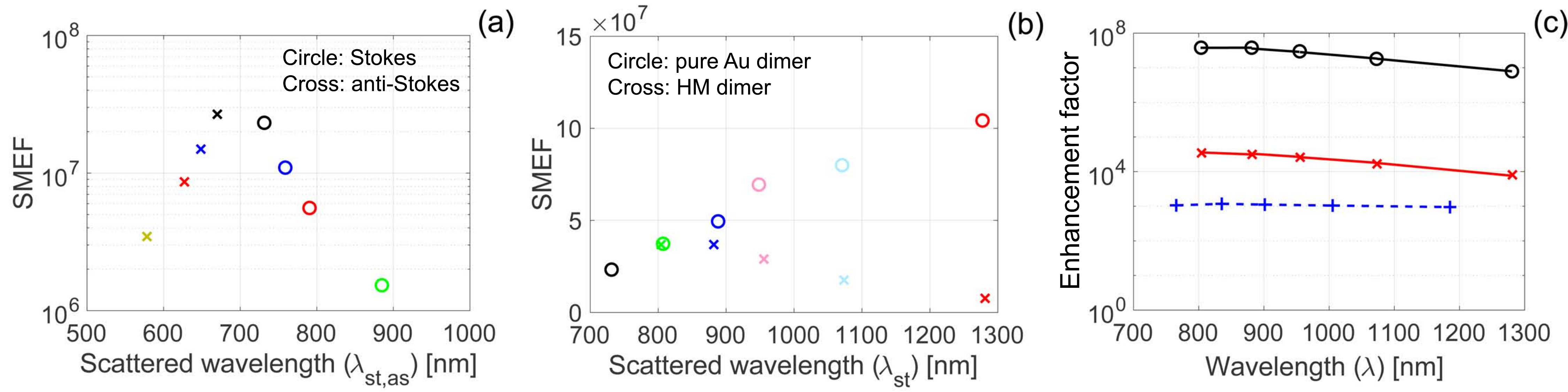}
        \caption{(a) Single molecule enhancement factor (SMEF) at center of dimer gap for both Stokes and anti-Stokes lines of different R6G Raman modes for Au nanorod dimer (95 nm nanorod length) as a function of the corresponding scattered wavelength (black: $\nu$ = 819 $\textnormal{cm}^{-1}$, blue: $\nu$ = 1290.5 $\textnormal{cm}^{-1}$, red: $\nu$ = 1559 $\textnormal{cm}^{-1}$, green: $\nu$ = 3000 $\textnormal{cm}^{-1}$) (b) SMEF of Stokes line of R6G Raman mode with $\nu$ = 819 $\textnormal{cm}^{-1}$ at  dimer gap center for Au and HM nanorod dimers, as a function of the corresponding scattered wavelength. Circles show Au nanorod dimers with nanorod lengths of 95 nm (black), 115 nm (green), 135 nm (blue), 150 nm (gray), 180 nm (cyan), and 230 nm (red). Crosses show HM nanorod dimers (95 nm nanorod length) with $f_m$ = 0.75 (green), 0.5 (blue), 0.4 (gray), 0.3 (cyan), and 0.2 (red). (c) Enhancement factors, including SMEF (Solid black - circle) and SMEF/$|\eta|^2$ (solid red - cross) of Stokes line of R6G Raman mode with $\nu$ = 819 $\textnormal{cm}^{-1}$ as a function of the corresponding scattered wavelength, and $|\eta|^2$ (dashed blue - plus) at center of nanorod dimer gap as a function of resonant wavelength $\lambda_\textnormal{LSP}$, for different HM nanorod dimers.}
        \label{fig:all_enhancements}
\end{figure*}

Next, we investigate the SERS enhancement for different HM and Au nanorod dimers, and assess how quenching affects the SERS performance in each case. As defined in Section \ref{Sec2}, the SERS performance of the different nanorod dimers can be assessed through the SMEF, here defined for a molecule located in the center of the gap region. To do this we model the fluorescent molecule Rhodamine 6G (R6G), and use a few of its Raman modes with different Raman shifts $\nu$ and the associated Raman activities $\textnormal{RA} = R_{nn}^2$ as examples. In particular, we consider three different R6G Raman modes at $\nu_1$ = 819 $\textnormal{cm}^{-1}$ ($\omega_m$ = 1.54 $\times 10^{14}$ rad/s  or 98.7 meV, $\textnormal{RA}$ = 6.2 $\mathring{\textnormal{A}}^4 \textnormal{amu}^{-1})$, $\nu_2$ = 1290.5 $\textnormal{cm}^{-1}$ ($\omega_m$ = 2.43 $\times 10^{14}$ rad/s  or 160.0 meV, $\textnormal{RA}$ = 5.9 $\mathring{\textnormal{A}}^4 \textnormal{amu}^{-1})$, and $\nu_3$ = 1559 $\textnormal{cm}^{-1}$ ($\omega_m$ = 2.94 $\times 10^{14}$ rad/s  or 193.5 meV, $\textnormal{RA}$ = 8.2 $\mathring{\textnormal{A}}^4 \textnormal{amu}^{-1})$ \cite{Watanabe2005}. In addition, the C-H vibrational Raman mode at $\nu_4$ = 3000 $\textnormal{cm}^{-1}$ ($\omega_m$ = 5.65 $\times 10^{14}$ rad/s  or 371.9 meV, $\textnormal{RA}$ = 7 $\mathring{\textnormal{A}}^4 \textnormal{amu}^{-1})$ is considered. The Raman scattered Stokes frequency is then given by $\omega_{st} = \omega_L - \omega_m$ and the anti-Stokes frequency would be $\omega_{\textnormal{as}} = \omega_L + \omega_m$. The corresponding Stokes and anti-Stokes scattered wavelengths are $\lambda_{\textnormal{st}} = 2\pi c/\omega_{\textnormal{st}}$ and $\lambda_{\textnormal{as}} = 2\pi c/\omega_{\textnormal{as}}$, respectively.

For our investigation, the excitation wavelength $\lambda_\textnormal{L}$ is set to the LSP wavelength $\lambda_{\textnormal{LSP}}$; e.g., using the Au nanorod dimer with nanorod length of 95 nm, $\lambda_\textnormal{L} = \lambda_{\textnormal{LSP}} = 699.7$ nm. The decay rate of the Raman modes $\gamma_m$ is taken to be 1.6 meV. The excitation wavelength affects not only the field enhancement factor $|\eta|$, but also determines the Stokes and anti-Stokes Raman scattered wavelengths, which would then dictate the overall SERS enhancement. The results shown here are based on the excitation intensity of 10 $\textnormal{mW}/\mu \textnormal{m}^2$, which is representative of typical laser power levels used in Raman spectroscopy systems \cite{Mak2013}.

The SMEF utilizing the Au nanorod dimer with nanorod length of 95 nm, for the example Raman modes (Stokes and anti-Stokes) as indicated by the respective scattered wavelengths $\lambda_\textnormal{st,as}$, is shown in Fig.\,\ref{fig:all_enhancements}a. The SMEF is higher for smaller Raman shifts (i.e., $\lambda_\textnormal{st,as}$ closer to $\lambda_\textnormal{LSP}$) and decreases significantly as the Raman shift increases. It can be seen that by moving the Raman shift of $\nu$ = 819 $\textnormal{cm}^{-1}$ to $\nu$ = 3000 $\textnormal{cm}^{-1}$, the SMEF decreases by approximately an order of magnitude for both the Stokes and anti-Stokes lines. In order to better observe the trend of SMEF as a function of $\lambda_\textnormal{LSP}$ of the nanorod dimer (either Au or HMs), we only present the results for the R6G Raman mode with $\nu$ = 819 $\textnormal{cm}^{-1}$, as shown in Fig.\,\ref{fig:all_enhancements}b. As the length of the Au nanorod dimer is increased to red-shift $\lambda_\textnormal{LSP}$ and thus the Raman scattered wavelength, the SMEF increases monotonically. In contrast, by decreasing $f_m$ of the HM nanorod dimer to red-shift the Raman scattered wavelength, the SMEF increases and then decreases again for lower $f_m$, with the maximum SMEF attained when $f_m$ is $\approx$ 0.6. Interestingly, the SMEF obtained by utilizing a HM nanorod dimer cannot exceed that of the Au dimer counterpart with corresponding $\lambda_\textnormal{LSP}$ for any $f_m$ values. While the employment of HMs can significantly increase the LDOS near nanostructures compared to pure metals such as Au (see Appendix \ref{appendix:Green_function}), the simultaneous rise in quenching proves detrimental to the process of SERS enhancement. As shown to be true also for the application to single-photon sources in Ref.~\onlinecite{Axelrod2017}, in which the $\beta$-factor (radiative quantum efficiency) decreases when HMs are utilized, it is presented here that SERS enhancement detected in the far-field of HM nanostructures is in general worse than by using the pure metal counterparts. 

\subsection{Delineating enhancement in excitation and scattering processes}

Besides the comparison of the SERS enhancement achievable by HM and pure metal nanostructures, another advantage of the approach undertaken in this work is that it can delineate the enhancement processes for the excitation and scattered fields, and rigorously determine  the contribution of each to the overall SERS enhancement. The contribution by the enhancement of the excitation field is captured by $|\eta|^2$, which is shown in Fig.\,\ref{fig:field_enhancement}a for the different nanorod dimers at their respective resonant wavelengths $\lambda_\textnormal{LSP}$. As such, the contribution by the enhancement of the scattered field is defined here by the quantity SMEF/$|\eta|^2$, which is shown in Fig.\,\ref{fig:all_enhancements}c along with values of $|\eta|^2$ and SMEF for different HM nanorod dimers at their respective resonant or scattered wavelengths in the same plot. It is observed that SMEF/$|\eta|^2$ (scattered field enhancement contribution to SERS) is in general over an order of magnitude higher than $|\eta|^2$ (excitation field enhancement contribution to SERS). This also elucidates why SERS enhancement as characterized by the SMEF is also over an order of magnitude higher than the conventional wisdom of the $(|\textbf{E}|/|\textbf{E}_0|)^4$ enhancement, which is equivalently given by $|\eta|^4 = (|\eta|^2)^2$. Note that this was also previously shown in Section \ref{Sec3}A, where the SERS enhancement using our approach is over an order of magnitude higher than simply $\textbf{E}$-field enhancement, with the experimental EF lying somewhere in between these theoretically estimated enhancement factors.

\begin{figure*}
        \centering
        \includegraphics[width=\textwidth]{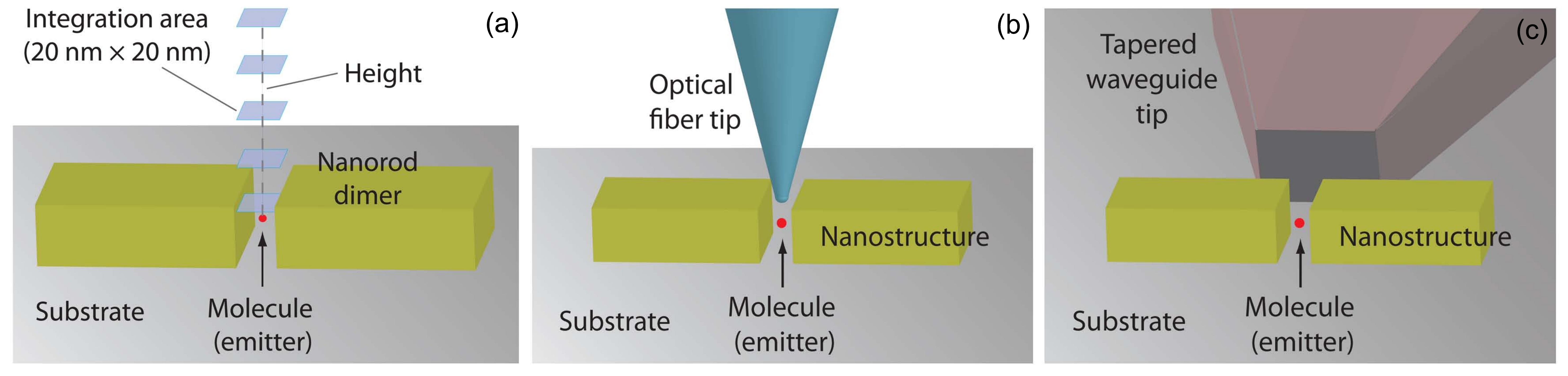}
        \caption{(a) Representation of the integration geometry for obtaining Raman scattered power, when the molecule of interest is in the vicinity of a plasmonic nanostructure in which the resonant mode is obtained from 3D Lumerical FDTD simulations, for the case of near-field probing. The height of the integration area is varied to obtain the distance (away from molecule) dependence of SERS enhancement. Schematics of the proposed near-field probed SERS (NFP-SERS) technique when implemented using (b) an optical fiber tip and (c) an integrated tapered waveguide tip.} 
        \label{fig:NPSERS}
\end{figure*}

As described in Ref.\,\onlinecite{Dezfouli2017}, the SERS enhancement is nonlinear with the pump field intensity, such that using a low pump on the order of 10's to 100's of $\textnormal{mW}/\mu\textnormal{m}^2$, the well-known $(|\textbf{E}|/|\textbf{E}_0|)^4$ enhancement rule is obtained based on the derivation in Section \ref{Sec2}; but upon high intensity excitation on the order of $10^5$ $\textnormal{W}/\mu\textnormal{m}^2$, a much stronger $(|\textbf{E}|/|\textbf{E}_0|)^8$ enhancement can be reached. The physical mechanism behind this nonlinearity is that an increase in the pump intensity increases the plasmonic-induced Raman scattering rate $J_{ph}$ [Eq.\,\eqref{eq:Jph}] beyond the intrinsic rate due to the thermal population contribution $\gamma_m\bar{n}^\textnormal{th}$. By comparing the value of $J_{ph}$ based on a set of excitation conditions, including the pump wavelength, the Raman shift and thus the Raman scattered wavelength, and intensity, to the value of $\gamma_m\bar{n}^\textnormal{th}$, it can be assessed whether the SERS device is operated in the linear or nonlinear regime. For example, consider the HM nanorod dimer with $f_m$ = 0.4 studied in this work; using a pump wavelength at $\lambda_\textnormal{LSP}$ = 901 nm with an intensity of 10 mW/$\mu\textnormal{m}^2$, the value of $J_{ph}$ for the Stokes mode with $\nu$ = 819 $\textnormal{cm}^{-1}$ is $\sim$4 $\textnormal{s}^{-1}$. Based on the excitation conditions used in this work, the value of $J_{ph}$ is approximately below 10 for all nanorod dimers studied. This is negligible in comparison to the value of $\gamma_m\bar{n}^\textnormal{th}$ $\approx$ 2.5 $\times$ $10^{12}$ $\textnormal{s}^{-1}$, which means that our results correspond to operation in the linear regime. As such, in this work we have specifically shown that even in the limit of linear pump regime, the $(|\textbf{E}|/|\textbf{E}_0|)^4$ SERS enhancement rule is strictly an approximation, but that the rigorously calculated SERS enhancement factor is in general significantly higher due to the higher contribution of the scattered field in comparison to the excitation field enhancement process.

\section{SERS in the near-field regime}
\label{Sec4}
In conventional Raman spectroscopy, the excitation of molecules and collection of Raman scattered light from the sample is performed through the use of an objective lens [Fig.\,\ref{fig:integration_geom}(a)], which means that far-field light is captured. This is true even when a SERS substrate, either on a planar surface \cite{Barcelo2012,Pieczonka2009,RodriguezLorenzo2009} or as plasmonic nanoparticles in solution \cite{Whitmore2011,Li1999,Liu2013}, is utilized to enhance the Raman scattered signal. This is the configuration for which the calculations of SERS enhancement thus far in this work has been applied to. The conclusion is that while the utilization of HMs in nanostructures can significantly increase the LDOS, quenching is also increased to the extent that the overall SERS enhancement in comparison to using pure metals is much reduced. However, there are also near-field probing schemes for capturing optical signals. One such technique is near-field scanning optical microscopy (NSOM) \cite{Dunn1999}, i.e., in which the nanometer sized tip of an optical fiber is placed close enough to the sample surface that only the evanescent field is transmitted and reflected before it is allowed to propagate into the far-field, thus capturing information regarding the surface at resolutions below the diffraction limit. In the realm of Raman spectroscopy, a nanoscale sharp metal tip can also be used to probe close to the surface of a sample surface for enhanced Raman signal from surface molecules, which is known as the technique of tip-enhanced Raman spectroscopy (TERS) \cite{Zhang2016}. With TERS, the metal tip acts as a plasmonic nanostructure that concentrates incident light intensity to interact with surface molecules, and causes both excitation light and scattered Raman signal to be enhanced, just like in the case of SERS with plasmonic nanostructures embedded on the substrate instead. However, the scheme involves excitation and collected light that propagate in the far-field, and thus it is not able to capture Raman enhanced signal in the near-field. 

In order to be able to combine the advantages of plasmonic enhancement and near-field collection of Raman signal, a new strategy must be developed. One idea would be to use a sharp optical fiber tip down to the nanometer scale to probe the near-field of plasmonic nanostructures on the surface of a substrate where the molecules under investigation are located. Another potential method is to fabricate a waveguide on the same substrate as the plasmonic nanostructure, with one end of the waveguide tapered to a sharp tip that sits very close to the hotspot of the plasmonic resonance, so that any scattered Raman signal can be collected in the near-field and directly routed to other on-chip components such as a detector. An additional advantage of this configuration is that the optical fiber tip can be used to locally excite only a single or a few plasmonic nanostructures evanescently, and thus lead to further Raman enhancement compared to the conventional SERS setup where a diffraction-limited beam excites many hotspots simultaneously. We term this new proposed technique as   ``Near-field Probed Surface Enhanced Raman Spectroscopy'' (NFP-SERS), which is schematically shown in Figs.\,\ref{fig:NPSERS}(b) and \ref{fig:NPSERS}(c), for the cases of the use of an optical fiber tip and an integrated waveguide probe, respectively.

\begin{figure*}
        \centering
        \includegraphics[width=\textwidth]{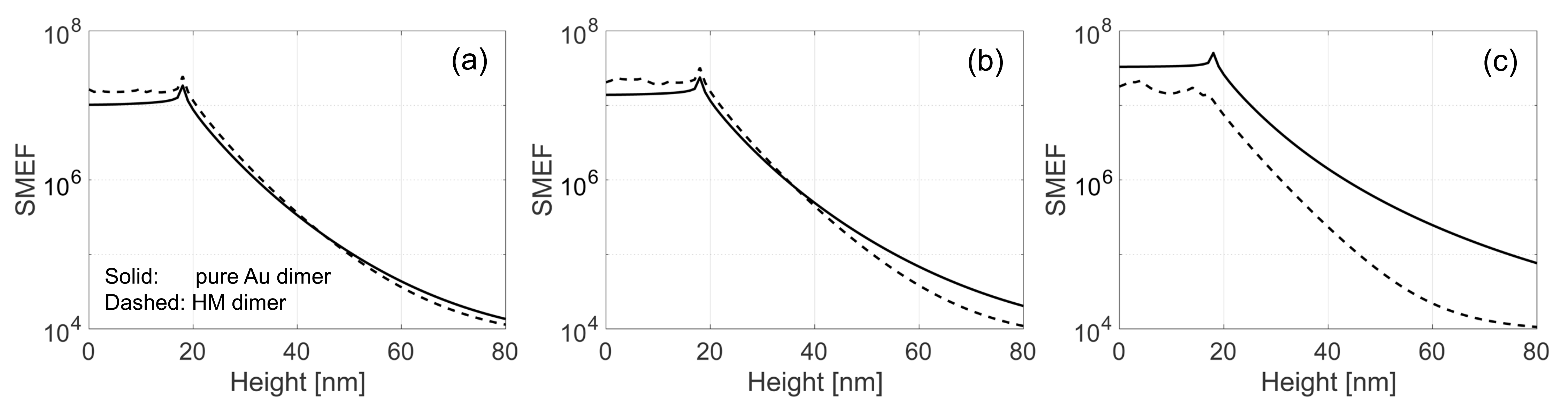}
        \caption{Single molecule enhancement factor (SMEF) detected by near-field probe as a function of vertical height away from molecule located at center of dimer gap for (a) $\textnormal{Au}$-$\textnormal{Si}_3\textnormal{N}_4$ HM $f_m$ = 0.75 nanorod dimer (95 nm nanorod length) and Au nanorod dimer (115 nm nanorod length), (b) $\textnormal{Au}$-$\textnormal{Si}_3\textnormal{N}_4$ HM $f_m$ = 0.5 nanorod dimer (95 nm nanorod length) and Au nanorod dimer (135 nm nanorod length), and (c) $\textnormal{Au}$-$\textnormal{Si}_3\textnormal{N}_4$ HM $f_m$ = 0.2 nanorod dimer (95 nm nanorod length) and Au nanorod dimer (230 nm nanorod length).} 
        \label{fig:SMEF_NF}
\end{figure*}

Here we show that the NFP-SERS can be modeled by the same approach we have described in this work without any additional complexity. Furthermore, it is discovered that HM based nanostructures can lead to higher SERS enhancement compared to using pure metals when the NFP-SERS configuration is implemented to capture Raman scattered light in the near-field. To investigate the Raman enhancement by NFP-SERS, all of the steps as previously described in Section \ref{Sec2} are followed, but the only difference is that the area over which the emitted Raman spectrum is integrated, would now be much smaller to represent the tip area of the optical fiber probe. As shown in Fig.\,\ref{fig:NPSERS}a, the size of the integration area is chosen to be 20 nm $\times$ 20 nm, so that it can even fit inside the gap region of the nanorod dimer, to facilitate probing of the hotspot down to a location directly next to the molecule at the center of the dimer gap. The Raman scattered power detected by this defined probe area is then calculated as a function of the vertical height of the probe away from the molecule at the center of the dimer gap. In Fig.~\ref{fig:SMEF_NF}, the near-field probed single molecule enhancement factors (SMEF) of the Stokes line of the R6G molecule with Raman shift of $\nu$ = 819 $\textnormal{cm}^{-1}$ for different HM and Au nanorod dimers are presented as a function of the height of the optical fiber probe tip away from the molecule. In each plot, the comparison is between a HM and an Au nanorod dimer with coinciding (or very nearby) resonant LSP wavelengths $\lambda_\textnormal{LSP}$; specifically, HM with $f_m$ = 0.75 is compared to Au nanorod dimer with nanorod length of 115 nm ($\lambda_\textnormal{LSP}$ $\approx$ 770 nm), then HM with $f_m$ = 0.5 is compared to Au nanorod dimer with nanorod length of 135 nm ($\lambda_\textnormal{LSP}$ $\approx$ 840 nm), and finally, HM with $f_m$ = 0.2 is compared to Au nanorod dimer with nanorod length of 230 nm ($\lambda_\textnormal{LSP}$ $\approx$ 1190 nm). For each of the cases of HM nanorod dimers with $f_m= 0.75$ and $f_m= 0.5$, it can be observed in Fig.\,\ref{fig:SMEF_NF}a and \ref{fig:SMEF_NF}b that the near-field detected SERS enhancement using HM can outperform the Au nanorod dimer counterpart when the optical fiber probe is less than 45 nm and 35 nm away from the molecule, respectively. This is despite  the fact that when detected in the far-field, the HM nanorod dimer attains a lower SMEF compared to the counterpart Au nanorod dimer for each of these cases ($\lambda_\textnormal{LSP}$ $\approx$ 770 nm and $\lambda_\textnormal{LSP}$ $\approx$ 840 nm), as shown in Fig.\,\ref{fig:all_enhancements}(b). For the case of the HM nanorod dimer with $f_m$ = 0.2 ($\lambda_\textnormal{LSP}$ $\approx$ 1190 nm), the SMEF cannot exceed that of the counterpart Au nanorod dimer with similar $\lambda_\textnormal{LSP}$ at any probe tip height away from the molecule, as deduced from Fig.\,\ref{fig:SMEF_NF}(c). When $f_m$ of the HM nanorod dimer is reduced to very low values, quenching is too strong such that the overall SERS enhancement even in the near-field is lower than for the pure metal implementation. 

From Fig.\,\ref{fig:SMEF_NF}(a) and \ref{fig:SMEF_NF}(b), it can be observed that the SERS enhancement when detected in the near-field of the HM nanorod dimer can exceed that of its pure metal counterpart. However, as the probe is moved away from the molecule at the center of the dimer gap where the plasmonic hotspot resides, the SMEF decays faster when HM is used compared to pure Au, and falls below that of when Au dimer is utilized for probe locations more than 50 nm away. This behavior is attributed to larger $\textbf{E}$-field enhancement but simultaneously higher quenching in HMs compared to in Au nanorod dimer; the enhancement is higher in the near-field for HMs, but propagation effects and quenching quickly reduce it to very low values after only a short distance. In Fig.\,\ref{fig:SMEF_NF}(c), it is seen that even in the near-field, the SMEF of the HM nanorod dimer ($f_m$ = 0.2) is much lower than its pure Au counterpart (nanorod length = 230 nm), because in this case the $\textbf{E}$-field enhancement $|\eta|$ for the HM dimer is actually lower than that of the Au dimer, due to very high quenching that serves to significantly reduce even the $\textbf{E}$-field intensity. The square of $\textbf{E}$-field enhancement for different Au and HM nanorod dimers have previously been shown in Fig.\,\ref{fig:field_enhancement}a, but here we specifically highlight the $\textbf{E}$-field enhancement factors for the dimers studied using near-field probing in Table \ref{table:field_enhancement_factors}.

\begin{table}
        
        \begin{tabular}{ | c | c | c |}
                \hline
                $\lambda_\textnormal{LSP}$ (approx.) [nm] & $|\eta|$ (Au dimer) & $|\eta|$ (HM dimer) \\ \hline
                770 & 28.8 & 32.6 \\ \hline
                840 & 32.3 & 34.3 \\ \hline
                1190 & 48.3 & 30.8 \\ \hline
        \end{tabular}
        \caption{Comparisons between the $\textbf{E}$-field enhancement factors $|\eta|$ of Au and HM nanorod dimers with coinciding (or very nearby) resonant LSP wavelengths $\lambda_\textnormal{LSP}$ $\approx$ 770 nm, 840 nm, and 1190 nm. Au dimer with nanorod length of 115 nm and HM dimer with $f_m$ = 0.75 have $\lambda_\textnormal{LSP}$ $\approx$ 770 nm, Au dimer with nanorod length of 135 nm and HM dimer with $f_m$ = 0.5 have $\lambda_\textnormal{LSP}$ $\approx$ 840 nm, and Au dimer with nanorod length of 230 nm and HM dimer with $f_m$ = 0.2 have $\lambda_\textnormal{LSP}$ $\approx$ 1190 nm.}
        \label{table:field_enhancement_factors}
        
\end{table}

Although for NFP-SERS, HMs can outperform pure metal nanostructures for Raman signal enhancement, the level of enhancement shown thus far in this work is quite low (i.e., less than 2 times). Nonetheless, within the niche of near-field SERS, we have shown that it is possible to produce higher Raman enhancement factors through the introduction of HMs, and thus based on presently available nano-fabrication capabilities, it is quite feasible that the engineering of new device structures in this direction may prove useful to further improve SERS performance.

\section{Discussion and Conclusions}
\label{Sec5}

We have explored the use of HM nanostructures to enhance SERS, and compared this to the performance of MNP resonators. Theoretically, we have used a first-principles framework based on Green functions and quantum optomechanical modeling of the interaction between a Raman active molecule and the nanoresonator, and it is used to calculate the SMEF, which has been shown to make predictions that are only $\sim$20 times higher (as a theoretical upper limit) than known experimentally determined SERS enhancement factors for some example gold nano-dimers, while predicting overall trends between different devices very well. Unlike previous work on theoretical calculations of SERS enhancement that typically under-estimates by 1 to 2-orders of magnitude, and thus attribute the discrepancy compared to experiment to chemical enhancement that has not been taken into account, here we can quantitatively determine the Raman enhancement based solely on the electromagnetic mechanism. The SMEF at the center of a representative Au nano-dimer with gap of 10 nm width can reach close to $10^8$. This work is partly motivated by the numerous studies in the literature reporting that hyperbolic materials exhibit much higher LDOS values compared to pure metals \cite{Cortes2012,Jacob2012}, but the implications of this effect for enhanced Raman scattering has largely been unexplored. Furthermore, there is also a pressing need to investigate new design strategies for improved SERS enhancement, with more reproducible and stable hotspots, which can notably benefit the efforts in developing single-molecule SERS \cite{Wang2013,Lee2013}.

We then\ investigated HM nanorod dimers constructed from a multilayer stack of Au and $\textnormal{Si}_3\textnormal{N}_4$ with varying metal filling fractions $f_m$, in conjunction with their counterpart pure Au nanorod dimers with coinciding resonant wavelengths $\lambda_\textnormal{LSP}$. The general conclusion is that for a given HM nanorod dimer with a certain $\lambda_\textnormal{LSP}$, the LDOS is significantly higher than its counterpart Au nanorod dimer. However, the concurrent increase in quenching leads to an overall lower SERS enhancement as detected. Moreover, as the $f_m$ is decreased, the SMEF first increases but eventually decreases again for lower $f_m$, such that there is a maximum SMEF at a certain $f_m$. For our choice of constituent materials Au and $\textnormal{Si}_3\textnormal{N}_4$, and nanorod dimer dimensions as shown in Fig.\,\ref{fig:Purcell_factor}, the SMEF is maximized for a $f_m$ between 0.5 and 0.75. Using HM nanorod dimers, the $\textbf{E}$-field intensity within the gap region where the molecule sits can be significantly increased as $f_m$ is reduced, but the simultaneous increase in $\textbf{E}$-field concentration inside the lossy dimer leads to higher absorption. Based on the theoretical approach in this work, the contribution to overall SERS enhancement can be clearly separated into the enhancement processes of the excitation and scattered fields. It is deduced that even in the linear pump regime by low excitation intensities, when the intrinsic Raman scattering rate is only due to thermal population of the molecule vibrational states, the contribution by scattered field enhancement is approximately an order of magnitude higher than that of the excitation ($\textbf{E}$-field) enhancement. This is a significant deviation from the well-known $(|\textbf{E}|/|\textbf{E}_0|)^4$ rule typically used to estimate SERS enhancement.

It is recognized that the nonlocal EM response of metals should be considered when nanostructures are investigated. This has not been taken into account in this work, but here we elucidate the effects that nonlocal response would have on the main results. For MNPs with characteristic size $<$10 nm, the pressure driven convective flow of charges causes a significant blue-shift in the LSP resonance, while charge carrier diffusion leads to further line broadening in addition to that caused by ohmic loss (quenching) \cite{Mortensen2014}. Specifically, the plasmonic modes that have induced positive and negative charges separated by a distance comparable to the length scale of the induced smearing of charge density away from the metal surface are affected by the nonlocal response \cite{Raza2015}. Although the sizes of the Au nanorod dimers studied (e.g., smallest one has dimensions of 35 nm $\times$ 35 nm $\times$ 95 nm) as well as the gap size (20 nm) are sufficiently large that nonlocal effects are negligible, the individual metal layers within the HM nanorod dimers have thicknesses of $<$10 nm, and thus nonlocal effects would be significant. For bulk HM, the addition of nonlocal response has been shown to lead to a cutoff of the dispersion for large wavevectors, which results in a finite maximal enhancement of the LDOS, as opposed to the theoretical infinite LDOS in the case of lossless materials and that both the sizes of the unit cell and emitter approach zero \cite{Yan2012}. For a single ultra-thin metal layer, as in those within the stratified HM dimers, the permittivity becomes anisotropic when nonlocal response is taken into account \cite{Laref2013}. In the out-of-plane direction (direction along the thickness), the plasma frequency increases with decreasing thickness \cite{Laref2013}, which is consistent with the trend that the blue-shift of the LSP resonance increases as the MNP size decreases \cite{Mortensen2014}. In the direction along the plane of the ultra-thin metal film (in-plane), the plasma frequency decreases as the thickness of the layer decreases, which is the opposite trend compared to in the out-of-plane direction \cite{Laref2013}. For each of the HM nanorod dimers investigated in this work, the dominant $\textbf{E}$-field of the LSP resonance is in the in-plane direction of the individual metal and dielectric layers (Fig.\,\ref{fig:Purcell_factor}), which means that by considering nonlocal effects, the plasma frequency of the metal layers would be decreased, and thus the LSP resonance of the HM dimer would also be further red-shifted in comparison to the results presented in this work. By decreasing the thickness of the metal layer from bulk to $<$10 nm, the imaginary part of the relative permittivity that is associated with loss is also increased appreciably when nonlocal effects are considered \cite{Laref2013}, both for the in-plane and out-of-plane directions. The combination of change in resonance wavelength, LDOS, and quenching of the HM nanorod dimer when nonlocal effects are included would modify the SMEF, but detailed calculations must be carried out in order to deduce the exact outcome. It should also be noted that, very recently, a QNM picture for the nonlocal regime of light confinement in small resonators has been introduced \cite{Dezfouli2017_2}, which extends the efficient performance of the semi-analytical techniques used in this work, to the regime of nonlocal behavior.

Although the general figures-of-merit are different, the overall conclusion that HMs are not as good at MNPs for SERS is in agreement with a recent study of using hyperbolic nanostructures for single photon emission enhancement, in which the quantum efficiency is shown to decrease in comparison to pure metal nanoresonators \cite{Axelrod2017}. The same conclusion can be stated for bulk hyperbolic materials in which the extraction of light from emitters in its vicinity into free space becomes less efficient in comparison to pure metals \cite{Galfsky2015,Sreekanth2014}. Although applications that require the detection of far-field light do not seem to benefit from the use of HMs, it has been shown that non-radiative dipole-dipole coupling, for example, can be much improved when HMs are implemented in comparison to pure metal plasmonic structures \cite{Biehs2016}. F{\"o}rster energy transfer is a non-radiative near-field resonant dipole-dipole interaction that is effective when the distance between donor and acceptor $d$ is on the order of 10 nm, and the strength of the process falls as $1/d^6$\,\,\cite{Andrew2004}. The use of a metal thin film that is placed between the donor and acceptor can increase the strength of the F{\"o}rster energy transfer process and thus increase the distance over which it can occur to be on the order of the wavelength (i.e., 100's of nanometers), because the energy transfer is mediated by the excitation of surface plasmon polariton (SPP) modes of the thin metal film \cite{Andrew2004}. For smaller donor-acceptor distances when SPP modes are not excited, then the rate of F{\"o}rster energy transfer close to the surface of a metal film is highly dependent on the spatial locations and orientations of the molecules, and either enhancement or reduction of the energy transfer rate can occur, which is not simply due to the enhanced LDOS \cite{Tumkur2015}. In Ref.\,\onlinecite{Biehs2016}, it is theoretically shown that the F{\"o}rster energy transfer between a donor-acceptor pair placed on each side of a HM slab mediated by SPP modes can actually be enhanced significantly compared to using a thin silver film of the same thickness. In this work, we show that by using a near-field probe to detect the Raman signal, an improvement is observed for the SERS enhancement by using hyperbolic stratified nanostructures, albeit only by a very small increase of less than 2 times. Nonetheless, it provides evidence for further work in this direction in the quest to increase SERS sensitivity, which is especially important for the burgeoning field of single-molecule Raman spectroscopy \cite{Wang2013,Lee2013} and other single-molecule detection techniques in general \cite{Arroyo2016,Anger2006}. 

Despite the negative message for HMs in the application to improve SERS enhancement, it is shown that the undesirable quenching effects can be reduced if detection of the Raman signal is performed in the near-field, such as by utilizing an optical fiber or an integrated waveguide tapered tip to probe in the vicinity of the plasmonic nanostructure where the molecule under investigation sits, which we term Near-field Probed SERS (NFP-SERS). It is observed that with near-field probing, the SERS enhancement for HM nanorod dimers can actually exceed their pure Au counterparts in some cases. For example, with the HM nanorod dimer with $f_m$ = 0.75, the SMEF detected in the near-field can surpass that obtained from its pure Au counterpart (nanorod length = 115 nm) when probed below 45 nm from the molecule at the center of the dimer gap, reaching close to a 2 times improvement.
Another route to potentially achieve higher SERS enhancement is through hybrid plasmonic dielectric type nanoresonators \cite{Yang2017}. Both of these strategies attempt to maintain high LDOS while circumventing the negative effect of quenching that limits the amount of detected scattered photons.

It must be recognized that although SERS is based on the scattering of photons, it is closely analogous to SE, in which there is both a non-radiative and a radiative part. As we have shown, only the Raman scattered photons that propagate to the detector are counted towards the SERS signal, whereas the remaining scattered photons are absorbed by the lossy metal or HM nanoresonator material. As such, one way to reduce quenching and thus improve Raman enhancement is to utilize lower loss plasmonic``metals" (negative permittivity materials); the search for novel low loss plasmonic materials is being pursued at a rapid rate \cite{Boltasseva2011,Khurgin2010,Gjerding2017}. With regards to HMs engineering, much work has already investigated in its bulk form how to improve the radiative efficiency of spontaneous emission into free space, such as with high index contrast gratings fabricated on the top surface \cite{Galfsky2015}. To achieve high LDOS and high radiative efficiency simultaneously, hybrid plasmonic dielectric nanoresonators have also been recently investigated \cite{Yang2017}, which has reported a Purcell factor of $>$$5000$ with quantum efficiency of $>$$90$$\%$ in the optical wavelengths inside a 2 nm gap between a dielectric nanoresonator and a metal substrate. This is in contrast to radiative efficiencies using pure metal nanoresonators that are typically $<$$70$$\%$ \cite{Axelrod2017}. Achieving high SERS enhancement requires both large LDOS and high efficiency of scattered photons that reaches the far-field, and thus the hybrid approach by combining the advantages of metals and dielectrics to construct nanostructures may prove fruitful for this endeavour. It represents a promising route going forward to utilize hyperbolic materials to further improve SERS.

\appendix

\section{QNM Green function}
\label{appendix:Green_function}

\begin{figure}
        \centering
        \includegraphics[width=\columnwidth]{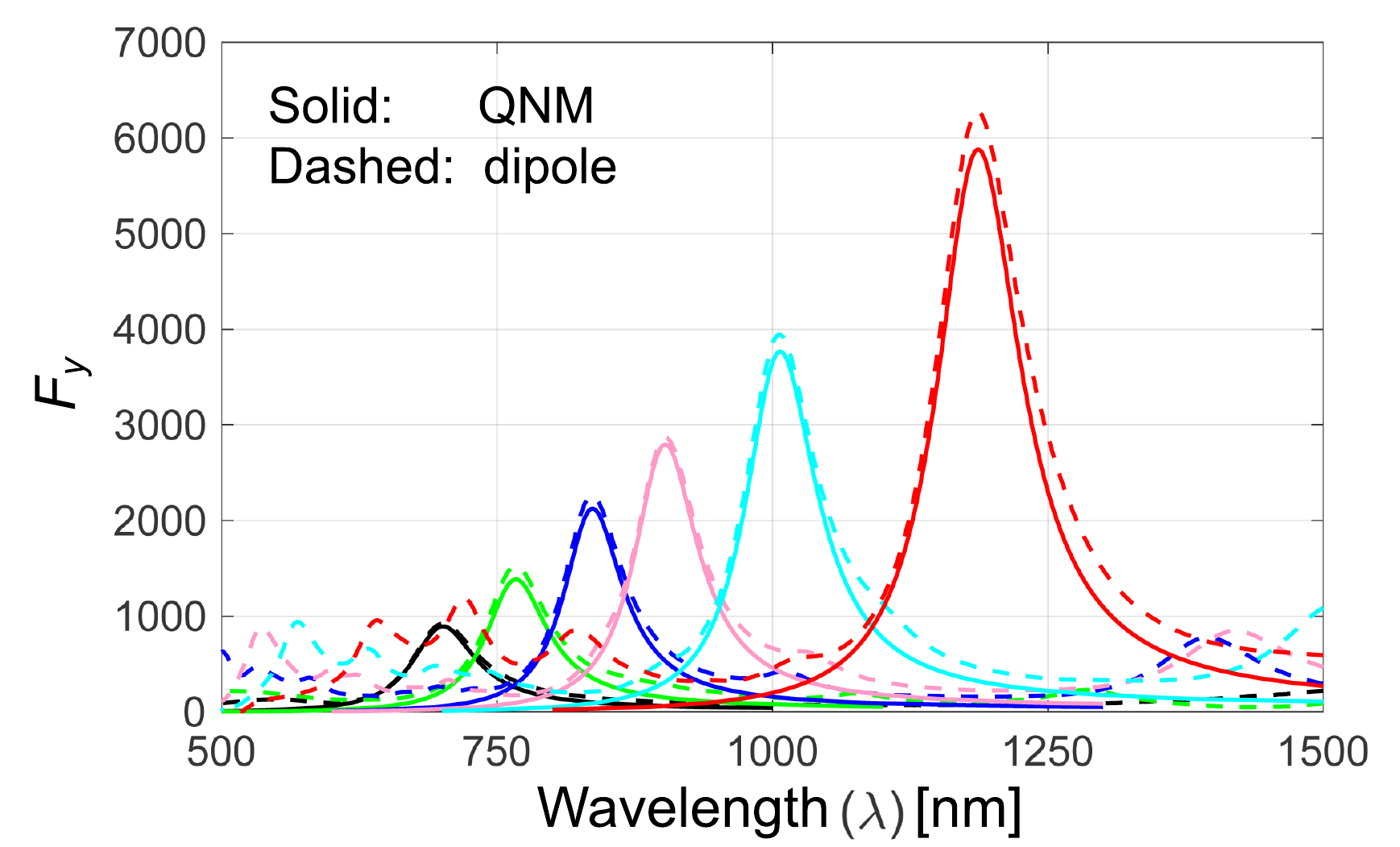}
        \caption{Comparison between QNM and full numerical calculation of Purcell factor for Au dimer with 95 nm nanorod length and different HM dimers, for a dipole emitter at the center of the dimer gap and polarized parallel to the dimer axis. Solid lines show results from using QNM theory, and dashed lines show results of full numerical (dipole) calculations. Au nanorod dimer (black), HM nanorod dimers (95 nm nanorod length) with $f_m$ = 0.75 (green), 0.5 (blue), 0.4 (gray), 0.3 (cyan), and 0.2 (red).}
        \label{fig:QNM_vs_dipole}
\end{figure}

Here we compare the corresponding Purcell factor of Eq.\,\eqref{eq:Purcell_factor} using the QNM Green function of Eq.\,\eqref{eq:Green_function}, against full dipole numerical calculations (i.e., with no approximations). Several devices using pure gold and HMs are used for such a compariosn as shown in Fig.\,\ref{fig:QNM_vs_dipole} (see caption for details). Note that the peak $F_{y}$ at $\lambda_\textnormal{LSP}$ from the QNM calculations is in general less than from dipole simulations, mostly due to other QNMs that are not taken into account. However, a very good degree of agreement, within 1\%-9\%, between QNM and fully numerical results is obtained. This is sufficient to allow us to explain the main physics of the LSP resonances, although the theory can also use the full dipole results as well.

\begin{figure}
        \centering
        \includegraphics[width=\columnwidth]{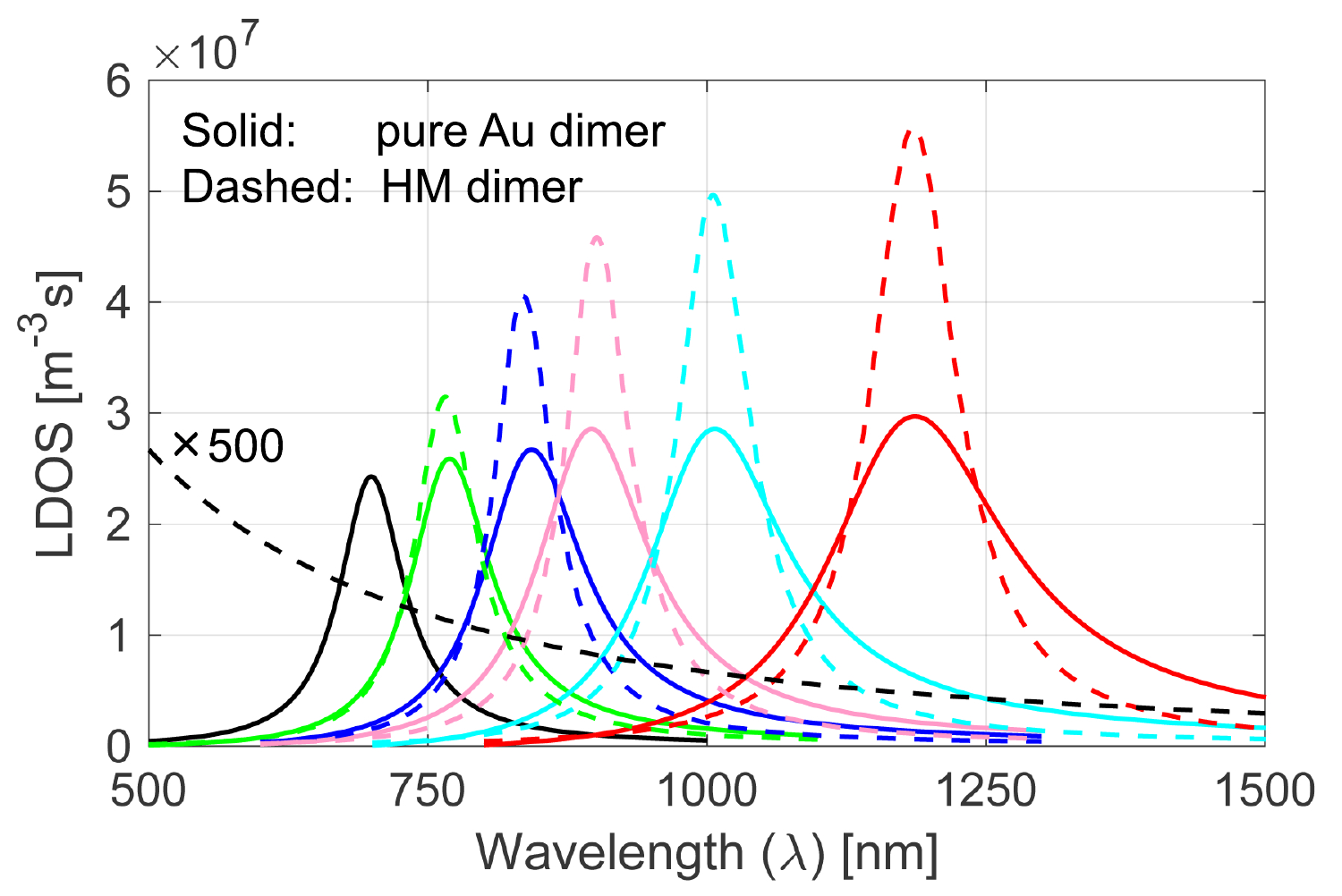}
        \caption{Photonic local density of states (LDOS) of Au and HM nanorod dimers for a dipole emitter at the center of the dimer gap and polarized parallel to the dimer axis. Solid lines show Au nanorod dimers with nanorod lengths of 95 nm (black), 115 nm (green), 135 nm (blue), 150 nm (gray), 180 nm (cyan), and 230 nm (red). Dashed lines show HM nanorod dimers (95 nm nanorod length) with $f_m$ = 0.75 (green), 0.5 (blue), 0.4 (gray), 0.3 (cyan), and 0.2 (red). Dashed black line shows the free space LDOS (multiplied by 500 times for graphical convenience).}
        \label{fig:Im_G}
\end{figure}

It is also worth looking at the LDOS, which is also obtained using QNM theory. The LDOS is directly proportional to the imaginary part of the Green function through \cite{Novotny2006}
\begin{equation}
\rho_n(\textbf{r}_0;\omega) = \frac{6}{\pi\omega}\textnormal{Im}\{\mathbf{n}\cdot\mathbf{G}(\mathbf{r}_0,\mathbf{r}_0;\omega)\cdot\mathbf{n}\}.
\end{equation}
For each of the Au and HM nanorod dimers investigated in this work, the LDOS is shown in Fig.\,\ref{fig:Im_G}. By increasing the length of the individual nanorods of the Au nanorod dimer from 95 nm to 230 nm to red-shift the peak $\lambda_\textnormal{LSP}$, the peak of the LDOS increases slightly from $\sim$2.43 $\times$ $10^7$ $\textnormal{m}^{-3}\textnormal{s}$ to $\sim$2.97 $\times$ $10^7$ $\textnormal{m}^{-3}\textnormal{s}$. However, the LDOS of the HM nanorod dimer is in general higher than that of the pure Au counterpart; it increases by almost a factor of 2 when $f_m$ decreases from 1 (pure Au) to 0.2, reaching a value of $\sim$5.58 $\times$ $10^7$ $\textnormal{m}^{-3}\textnormal{s}$.

\section{Details of the integration geometries used for SERS calculations}
\label{appendix:Integration_geom}

The SERS power $P_{\textnormal{SERS}}^{\textnormal{SM}}$ is calculated through first integration over a computational square surface representing the detector area located above the sample substrate, as shown in Fig.\,\ref{fig:integration_geom}b, which is followed by integration over the range of frequencies over which the Raman line covers. The integration over the flat surface of a square is performed because the Green function of the plasmonic nanorod dimer is calculated using data from Lumerical FDTD \cite{Lumerical}, which is based on a cartesian coordinate system. In a practical experiment, the Raman scattered light is first collected by an objective lens which is located above the sample, and thus the detection area can be effectively considered to be the area of the entrance pupil on the objective lens over which light can be collected (Fig.\,\ref{fig:integration_geom}a). To determine the area of integration given a fixed height $h$ away from the molecule, the angle $\theta$ must first be calculated based on knowledge of the numerical aperture (NA) of the objective lens employed to focus excitation light onto the sample and collect Raman scattered light, which is given by $\textnormal{NA} = n\sin\theta$. Once $\theta$ is known, then the side length $L$ of the square surface of integration can be determined (Fig.\,\ref{fig:integration_geom}b). Here, the height $h$ used in the calculation is the vertical distance from the molecule location to the boundary of the simulation region (on the order of 100's of nanometers as limited by reasonable size for the simulation region), which is practically different from the height of the objective lens in an actual experiment (on the order of millimeters as dictated by the objective lens' working distance). Nonetheless, the total Raman scattered power spectral density captured is equivalent because the integration area used for the calculation is correctly scaled down (i.e., the collection angle $\theta$ is constant). To calculate the free space Raman power $P_{\textnormal{RS}}^{\textnormal{SM}}$, however, the surface integration is performed over the portion of the surface of a sphere (spherical cap) whose base circle has equivalent area as the square integration area previously defined, which is schematically shown in Fig.\,\ref{fig:integration_geom}c. The integration is over a spherical surface because the formulation of the free space Green function is analytical and only depends on the radial position vector $\textbf{r}$ with respect to the molecule location \cite{Novotny2006}. First, the square integration area $A$ is used to calculate the effective radius of the base circle through $r_{\textnormal{eff}} = \sqrt{A/\pi}$, and then the radius of the spherical cap $r$ can be determined by $r = r_{\textnormal{eff}}/\sin\theta$. Finally, the spherical cap surface area of integration is $A_{\textnormal{sph.\,cap.}} = 2\pi r^2(1 - \cos\theta)$.

\section{Details of FDTD simulation and geometrical parameters of nanorod dimers}
\label{appendix:sim_setup}

In this Appendix, we summarize  the relevant geometrical parameters used in the design of different nanorod dimer devices, as well as the various FDTD parameters used for our numerical simulations. The metal filling fractions for the HM nanorod dimers investigated are $f_m = 1$ (pure metal), 0.75, 0.5, 0.4, 0.3, and 0.2. The dimensions of the HM nanorod dimers are kept constant and the same as the reference Au nanorod dimer (Fig.\,\ref{fig:Purcell_factor}). The effective medium approximation is not used, but rather the HM is constructed from a stratified metal-dielectric stack with ultra-subwavelength layer thicknesses. For each of the dimers with a given $f_m$, the total thickness of one period of metal-dielectric bilayer is maintained at 10 nm while the individual Au and $\textnormal{Si}_3\textnormal{N}_4$ layer thicknesses vary depending on the $f_m$. For example, the HM nanorod dimer with $f_m = 0.5$ would consist of Au and $\textnormal{Si}_3\textnormal{N}_4$ layers each 5 nm thick; however, for $f_m = 0.75$, the Au layer thickness would be 7.5 nm while the $\textnormal{Si}_3\textnormal{N}_4$ layer is 2.5 nm thick. The top layer is set to be Au, and the bottom layer would have a thickness that keeps the total height of the nanorod to be 35 nm (i.e., the bottom layer of the $f_m$ = 0.75 nanorod is Au with a thickness of 5 nm). The Au nanorod dimers have lengths of 115 nm, 135 nm, 150 nm, 180 nm, and 230 nm, with $\lambda_\textnormal{LSP}$ corresponding to those of HM nanorod dimers with $f_m$ = 0.75, 0.5, 0.4, 0.3, and 0.2, respectively.

The stratified HM nanorod dimers described in this work can be fabricated by a combination of high-resolution electron-beam lithography (EBL) to define the lateral pattern of the dimers, electron-beam evaporation to deposit the alternating metal and dielectric multilayer stack, and a lift-off process to remove excess materials outside of the dimer structure. In fact, stratified HM nanostructures have been fabricated as presented in Ref.\,\citenum{Yang2012}, and details of the fabrication process can be borrowed and applied for the nanorod dimers investigated here. Although the requirement of ultra-thin films proposed in this work (i.e., down to 2 nm) would be problematic for electron-beam evaporation, as continuous metal films are required to be $>$$6.5$ nm thick \cite{Hovel2011}, below which the films become granular and form islands, epitaxial techniques can produce ultra-thin layers with atomically flat interfaces and may be further explored for the fabrication of stratified HMs \cite{Ayers2007,Sands1990,Balakrishnan2005}. 

The Lumerical FDTD simulation domain size is 1.5 $\times$ 1.5 $\times$ 1.5 $\mu\textnormal{m}$, and a conformal meshing scheme with a maximum mesh step size of 40 nm in all directions and a smaller refined mesh of 1 nm around the nanorod dimer is used. In each direction, 20 perfectly matching layers with symmetric (antisymmetric) boundary condition in the $\textit{z}$ ($\textit{y}$) direction are employed. The time step of the simulation is 1.9066 as.

\bibliography{HMK_Wong_PRB_Aug_2017_ref_revised_2}

\end{document}